\documentclass[aps,prd,onecolumn,groupedaddress,showpacs,nofootinbib,amssymb
]{revtex4}
\usepackage[dvips]{graphicx}
\usepackage{amssymb}
\usepackage{amsmath}
\usepackage{graphicx}
\usepackage{amsfonts}
\usepackage{bm}
\begin{document}

\title{Dynamical Systems Perspective of Cosmological Finite-time Singularities in $f(R)$ Gravity and Interacting Multifluid Cosmology}
\author{S.D. Odintsov,$^{1,2,3}$\,\thanks{odintsov@ieec.uab.es}
V.K. Oikonomou,$^{4,5,6}$\,\thanks{v.k.oikonomou1979@gmail.com}}
\affiliation{$^{1)}$ ICREA, Passeig Luis Companys, 23, 08010 Barcelona, Spain\\
$^{2)}$ Institute of Space Sciences (ICE,CSIC) C. Can Magrans s/n,
08193 Barcelona, Spain\\
$^{3)}$ Institute of Space Sciences of Catalonia (IEEC),
Barcelona, Spain\\
$^{4)}$ Department of Physics, Aristotle University of Thessaloniki, Thessaloniki 54124, Greece\\
$^{5)}$ Laboratory for Theoretical Cosmology, Tomsk State University
of Control Systems
and Radioelectronics (TUSUR), 634050 Tomsk, Russia\\
$^{6)}$ Tomsk State Pedagogical University, 634061 Tomsk, Russia
}

\tolerance=5000

\begin{abstract}
In this work we shall investigate the occurrence of future cosmological finite-time singularities in the dynamical system corresponding to two cosmological theories, namely that of vacuum $f(R)$ gravity and that of three fluids. As we shall make clear, a finite-time cosmological singularity may be entirely different from a finite-time singularity of a dynamical system, since the latter mainly depends on the behavior of the dynamical system variables. The vacuum $f(R)$ gravity is an example for which the variables we will choose to quantify the phase space dynamics, do not necessarily blow-up near a cosmological singularity. After appropriately choosing the variables, we shall investigate the behavior of the corresponding dynamical system near some types of cosmological finite-time singularities, for some limiting cases in which we can produce analytic solutions for the dynamical variables. The most interesting case from both a mathematical and physical point of view, is the Big Rip case, and particularly in the limiting case of a very strong singularity. The physically appealing outcome is that the resulting non-autonomous dynamical system is attracted asymptotically to an accelerating attractor solution, with equation of state parameter $w_{eff}=-1$. Our analytic results, show that an extremely strong Big Rip singularity in vacuum $f(R)$ gravity theories is always related to an accelerating solution, or tends to acceleration. The converse statement though may not be true. We also perform the same analysis for the Type IV finite-time singularity, and we investigate the behavior of the dynamical system near the Type IV singularity, in the case that the singularity is extremely soft, in which case we are able to produce analytic expressions for the dynamical solutions. Also we briefly discuss how the removal of the finite-time singularity may be achieved by the addition of an $R^2$ term in the $f(R)$ gravity action. The second cosmology we shall study is a multifluid cosmology, consisting of three fluids, the interacting dark matter and dark energy fluids, and the baryonic fluid. By appropriately choosing the variables, we will show that the dynamical system can become an autonomous polynomial dynamical system, in which case, by using a dominant balance analysis, we shall investigate the occurrence of finite-time singularities in this system. We also study numerically and analytically, in some detail, the phase space of the dynamical system for some specific forms of the dark energy-dark matter interaction term.
\end{abstract}

%PACS numbers: 04.50.Kd, 95.36.+x, 98.80.-k, 98.80.Cq
%\pacs{04.50.Kd, 95.36.+x, 98.80.-k, 98.80.Cq,11.25.-w}

\maketitle

\section{Introduction}

The dark sector of the Universe controls the current evolution, with the luminous matter being only nearly the $4\%$ of the total energy density of the Universe. Hence, the main mechanism that drives the current evolution, is still unknown, although it overwhelms the dynamics. The dark sector of the Universe consists of the dark energy, which is responsible for the late-time acceleration of the Universe, which was observed in the late 90's \cite{Riess:1998cb}, and of the dark matter. The latter was believed for many years to be quantified in terms of a non-interacting particle, and many proposals exist in the literature, regarding particle dark matter candidates \cite{Oikonomou:2006mh}. However, no experiment has ever verified the particle nature of dark matter, thus the problem of dark matter still remains unsolved.

With regard to dark energy, the most successful description is the $\Lambda$CDM model, which is quite robust against the observational data up to date. However there exist various alternative theoretical proposals that may also describe an accelerating late-time evolution, with the most appealing of these alternative scenarios, being the $f(R)$ gravity description, for reviews see \cite{reviews1,reviews2,reviews3,reviews4,reviews5,reviews6}. Actually, it is possible in the context of $f(R)$ gravity to describe both the early and the late-time acceleration eras in a unified way \cite{Nojiri:2003ft}. Thus the focus of theoretical cosmology at present day is  on the late-time and early-time accelerating eras and also on dark matter. Many descriptions may in principle describe successfully some aspects of the above eras, but no complete answer is given up to date.

Due to the fact that the dark sector of our Universe is unknown, it is possible that the dark energy and dark matter sectors are described by interacting fluids, viscous or not. Actually, the fluid approach in cosmology has been adopted in the literature in order to produce viable cosmological evolutions, see for example \cite{Barrow:1994nx,Tsagas:1998jm,HipolitoRicaldi:2009je,Gorini:2005nw,Kremer:2003vs,Carturan:2002si,Buchert:2001sa,Hwang:2001fb,Cruz:2011zza,Oikonomou:2017mlk,Brevik:2017msy,Nojiri:2005sr,Capozziello:2006dj,Nojiri:2006zh,Elizalde:2009gx,Elizalde:2017dmu,Brevik:2016kuy,Balakin:2012ee,Zimdahl:1998rx,Brevik:2018azs}. An interaction in between the dark sector components, is supported by the fact that dark energy dominates over dark matter after galaxy formation. In addition, it is known that dark energy and dark matter strongly depend on each other, and actually, due to the degeneracy of dark energy models, $\Omega_m$ cannot be measured \cite{Kunz:2007rk}. There exist various proposals in the literature that study interacting dark energy-dark matter fluid models, see for example \cite{Gondolo:2002fh,Farrar:2003uw,Cai:2004dk,Bamba:2012cp,Guo:2004xx,Wang:2006qw,Bertolami:2007zm,He:2008tn,Valiviita:2008iv,Jackson:2009mz,Jamil:2009eb,He:2010im,Bolotin:2013jpa,Costa:2013sva,Boehmer:2008av,Li:2010ju,Yang:2017zjs} and references therein. However, one should be cautious with these models, due to the fact that interacting dark energy-dark matter models can suffer from instabilities, as it was shown for example in \cite{Valiviita:2008iv}. Indeed, a simple interaction term between the dark sector fluids, with a simple constant equation of state parameter for the dark energy sector, causes instabilities to the early times dark sector perturbations, since the curvature perturbation blows up on super-Hubble scales, see \cite{Valiviita:2008iv} for details on this. In addition, the growth of matter perturbations may be directly affected by a non-trivial interaction between the dark sector components. In all the cases, baryons cannot be coupled with the dark energy and dark matter fluids, since this would lead to a fifth force in the Universe, which is an unwanted feature, at least for the moment.

The discovery of the dark energy epoch of the Universe made compelling the effective fluid description of our Universe, with a negative equation of state parameter. As a result it was discovered that the dark energy Universe may be in quintessence or phantom- regime currently or in the near future. Then, it is quite possible that the future Universe develops some finite-time future singularity (for more detail, see section III). Despite many efforts, the physical understanding of the quantitative features of the Universe while approaching this finite-time singularity are not well understood up to date. To this end, in this paper we shall focus in two major problems, related with finite-time singularities, in the context of $f(R)$ gravity and coupled models of dark matter and dark energy, by using a multifluid approach. In both cases we shall use the dynamical systems approach and we shall study the occurrence of finite-time singularities in cosmological systems, from a dynamical systems perspective. The dynamical systems approach in cosmology  has many attributes, since the phase space can provide useful information about the cosmological solutions, in terms of the fixed points of the dynamical system, and about the stability of the cosmological solutions. In the literature, many studies use the dynamical systems approach in cosmology \cite{Boehmer:2014vea,Bohmer:2010re,Goheer:2007wu,Leon:2014yua,Leon:2010pu,deSouza:2007zpn,Giacomini:2017yuk,Kofinas:2014aka,Leon:2012mt,Gonzalez:2006cj,Alho:2016gzi,Biswas:2015cva,Muller:2014qja,Mirza:2014nfa,Rippl:1995bg,Ivanov:2011vy,Khurshudyan:2016qox,Boko:2016mwr,Odintsov:2017icc,Odintsov:2017tbc,Oikonomou:2017ppp,Ganiou:2018dta,Shah:2018qkh,Chakraborty:2018bxh,Bari:2018edl},
see also \cite{Odintsov:2015wwp}, and for a recent informative review see \cite{Bahamonde:2017ize}. Recently, we investigated the $f(R)$ gravity dynamical system, focusing on the early-time behavior of the phase space, and we found useful information, regarding the de Sitter attractors of the cosmological system \cite{Odintsov:2017tbc}. The first part of this study is devoted on the behavior of the vacuum $f(R)$ gravity behavior near all types of cosmological finite-time singularities. Specifically, we shall be interested in the behavior of the phase space near a Big Rip or near a Type II, III and IV singularities. To this end, we shall use the formalism we developed in Ref. \cite{Odintsov:2017tbc}, and we shall investigate how the dynamical system becomes near each one of the aforementioned finite-time singularities. Our findings are particularly interesting, and the most sound outcome is the fact that a Big Rip singularity is always connected with an accelerating attractor, when the singularity is extremely strong, which is a limiting case of the free parameters as we show. By using the formalism we developed in \cite{Odintsov:2017tbc}, we shall demonstrate that the vacuum $f(R)$ gravity dynamical system near the Big Rip is a non-autonomous dynamical system, which is formed in such a way so that it can be integrated analytically. Thus we examine the behavior of the solutions near the Big Rip, and as we demonstrate, the dynamical evolution ends up to an accelerating attractor, which is stable and also it corresponds to the attractor of the asymptotically autonomous dynamical system. The latter feature has also some mathematical implications, which we briefly mention without getting into much details. Also, we investigate which $f(R)$ may generate the cosmological evolution near the Big Rip singularity, directly from the phase space behavior, which is a new approach in the field of $f(R)$ gravity studies. After discussing the Big Rip case, we also discuss in some detail the behavior of the vacuum $f(R)$ gravity near Type II, III and IV singularities. As we show, analytic results can be obtained only for the case of a Type IV singularity, in the limit that this is extremely soft, so this is a limiting case of the free parameters too. So for the Type IV case, we find the behavior of the phase space near the singularity, and also the $f(R)$ gravity which realizes locally near the singularity, such a behavior. As we will show, the dynamical variables do not necessarily blow up at the finite-time singularity, and this is a very important feature that must be noted. So a physical singularity may not render singular the dynamical variables used, or in the converse way of thinking, a blow up of a dynamical variable in the phase space, may not necessarily indicate a physical finite-time singularity of the dynamical system. This strongly depends on the functional form of the dynamical variables, so this brings us to the second part of this work. Particularly, in the second part of this work, we shall investigate the finite-time singularity structure of a multifluid Universe, consisting from interacting dark energy-dark matter fluids and from baryonic matter, which does not interact with the other fluids. We shall construct the dynamical system corresponding to the cosmological system of the three fluids in such a way so that it is an autonomous polynomial dynamical system. The condition that restricts the dynamical system to be of polynomial form is crucial, since the whole singularity analysis we shall perform is based exactly on this polynomial condition. The finite-time singularity analysis for this kind of polynomial autonomous dynamical systems was performed in Ref. \cite{goriely}, and later on was further analyzed in the context of cosmological singularities in Ref. \cite{barrowcotsakis}. By using the results of the theorem developed in \cite{goriely}, we shall perform a detailed dominant balance analysis of the multifluid dynamical system. We shall discuss the properties of a potential physical finite-time singularity and the relation it has with a finite-time singularity in a variable of the dynamical system and we shall examine the conditions that must hold true in order for a dynamical system finite-time singularity to be an actual physical singularity of the cosmological system. Our analysis indicates that the resulting dynamical system does not have general solutions (no global attractors) which may lead the dynamics to a finite-time singularity. Finally, we perform an analytic and numerical study of the phase space of the resulting dynamical system, in order to see the behavior of the trajectories and also to find the fixed points of the dynamical system. As we demonstrate, some de Sitter vacua are fixed points, and these depend strongly on the interaction between dark energy and dark matter.

This paper is organized as follows: In section II we present the general form of the vacuum $f(R)$ gravity, and we review the case of the de Sitter attractor fixed point. This will be needed in the Big Rip study. In section III we perform the finite-time singularity analysis of the vacuum $f(R)$ gravity dynamical system in full detail. After discussing in brief the essential features of finite-time cosmological singularities, we divide our analysis in two subsections focusing on the Big Rip case and on the rest of the types of singularities. As we shall demonstrate, the Big Rip case has appealing mathematical and physical implications. In the end of section III, we discuss in brief the difficulties one confronts when the study is performed for the $f(R)$ gravity with radiation and non-relativistic matter included. In section IV we study the finite-time singularities structure of the dynamical system corresponding to the cosmological system of the dark energy, dark matter and baryonic fluid. We shall demonstrate how to choose the dynamical variables in order for the dynamical system to be an autonomous polynomial dynamical system and we perform a dominant balances analysis of the resulting dynamical system in order to reveal if any physical finite-time singularities occur. We also discuss in detail the conditions which need to hold true in order for a finite-time singularity of the dynamical system, to be an actual cosmological finite-time singularity of the cosmological system. Finally, the conclusions follow in the end of the paper.

Before starting our presentation, we shall describe the background geometry which shall be assumed in this work. Particularly, we shall use a flat Friedmann-Robertson-Walker (FRW) metric, with line element,
\begin{equation}\label{frw}
ds^2 = - dt^2 + a(t)^2 \sum_{i=1,2,3} \left(dx^i\right)^2\, ,
\end{equation}
with $a(t)$ being the scale factor. In addition, the Ricci scalar for the flat FRW metric reads,
\begin{equation}\label{ricciscalaranalytic}
R=6\left (\dot{H}+2H^2 \right )\, ,
\end{equation}
with $H=\frac{\dot{a}}{a}$, being the Hubble rate.

\section{The Vacuum $f(R)$ Gravity Dynamical System}

In a recent work, it was demonstrated that the $f(R)$ gravity equations of motion can be formed in such a way so that the resulting dynamical system is rendered autonomous in some cases \cite{Odintsov:2017tbc}. We shall use the theoretical framework of Ref. \cite{Odintsov:2017tbc}, since for the study of finite time singularities, the resulting dynamical system can have an elegant form as we shall show shortly. However, the obtained dynamical system is not autonomous in the case at hand, however it can be integrated analytically in some limiting cases, which is a great advantage in comparison with other approaches on $f(R)$ gravity existing in the literature. Let us recall how the $f(R)$ gravity dynamical system is formed, and for details we refer the reader to Ref. \cite{Odintsov:2017tbc}. We start off with the vacuum $f(R)$ action, which has the following form,
\begin{equation}\label{action}
\mathcal{S}=\frac{1}{2\kappa^2}\int \mathrm{d}^4x\sqrt{-g}f(R)\, ,
\end{equation}
where $\kappa^2=8\pi G=\frac{1}{M_p^2}$, $G$ is the gravitational constant,  and in addition, $M_p$ is the Planck
mass scale. By employing the metric formalism approach, the gravitational equations of motion can be obtained by varying the $f(R)$ gravity action (\ref{action}) with respect to the metric $g_{\mu \nu}$, and we obtain the following gravitational equations,
\begin{equation}\label{eqnmotion}
F(R)R_{\mu \nu}(g)-\frac{1}{2}f(R)g_{\mu
\nu}-\nabla_{\mu}\nabla_{\nu}f(R)+g_{\mu \nu}\square F(R)=0\, ,
\end{equation}
which can be further cast in the following way,
\begin{align}\label{modifiedeinsteineqns}
R_{\mu \nu}-\frac{1}{2}Rg_{\mu
\nu}=\frac{\kappa^2}{F(R)}\Big{(}T_{\mu
\nu}+\frac{1}{\kappa^2}\Big{(}\frac{f(R)-RF(R)}{2}g_{\mu
\nu}+\nabla_{\mu}\nabla_{\nu}F(R)-g_{\mu \nu}\square
F(R)\Big{)}\Big{)}\, ,
\end{align}
where $F(R)=\frac{\partial f(R)}{\partial R}$. By using the FRW metric of Eq. (\ref{frw}), the cosmological equations take the following form,
\begin{align}
\label{JGRG15} 0 =& -\frac{f(R)}{2} + 3\left(H^2 + \dot H\right)
F(R) - 18 \left( 4H^2 \dot H + H \ddot H\right) F'(R)\, ,\\
\label{Cr4b} 0 =& \frac{f(R)}{2} - \left(\dot H + 3H^2\right)F(R) +
6 \left( 8H^2 \dot H + 4 {\dot H}^2 + 6 H \ddot H + \dddot H\right)
F'(R) + 36\left( 4H\dot H + \ddot H\right)^2 F'(R) \, ,
\end{align}
with $F(R)=\frac{\partial f}{\partial R}$, $F'(R)=\frac{\partial^2
F}{\partial R^2}$, and $F''(R)=\frac{\partial^3 F}{\partial R^3}$. In order to obtain the $f(R)$ gravity cosmological dynamical system that corresponds to the cosmological equations (\ref{JGRG15}), we introduce the following dimensionless variables $x_1$, $x_2$ and $x_3$,
\begin{equation}\label{variablesslowdown}
x_1=-\frac{\dot{F}(R)}{F(R)H},\,\,\,x_2=-\frac{f(R)}{6F(R)H^2},\,\,\,x_3=
\frac{R}{6H^2}\, .
\end{equation}
For the purposes of this article, it is more convenient to use the $e$-foldings number $N$ as a dynamical variable, instead of the cosmic time, so we shall use the following differentiation rule,
\begin{equation}\label{specialderivative}
\frac{\mathrm{d}}{\mathrm{d}N}=\frac{1}{H}\frac{\mathrm{d}}{\mathrm{d}t}\,
,
\end{equation}
By using the variables $x_1$, $x_2$ and $x_3$ given in Eq. (\ref{variablesslowdown}), in combination with the gravitational equations of motion (\ref{JGRG15}), the following dynamical system is obtained for the vacuum $f(R)$ gravity,
\begin{align}\label{dynamicalsystemmain}
& \frac{\mathrm{d}x_1}{\mathrm{d}N}=-4-3x_1+2x_3-x_1x_3+x_1^2\, ,
\\ \notag &
\frac{\mathrm{d}x_2}{\mathrm{d}N}=8+m-4x_3+x_2x_1-2x_2x_3+4x_2 \, ,\\
\notag & \frac{\mathrm{d}x_3}{\mathrm{d}N}=-8-m+8x_3-2x_3^2 \, ,
\end{align}
where we introduced the parameter $m$ which is equal to,
\begin{equation}\label{parameterm}
m=-\frac{\ddot{H}}{H^3}\, .
\end{equation}
The parameter $m$ is the only term in the dynamical system (\ref{dynamicalsystemmain}), which contains an explicit time-dependence, or equivalently $N$-dependence for a general functional form of the Hubble rate. In Ref. \cite{Odintsov:2017tbc}, the cases for which $m=$const were studied. For example, the case $m=0$ can be produced by an exact de Sitter scale factor, that is $a(t)=e^{\Lambda t}$, where $\Lambda=$constant, or from a quasi de Sitter scale factor $a(t)=e^{H_0t-H_it^2}$, where $H_0,H_i$ are constants. Also as was shown in \cite{Odintsov:2017tbc}, the case $m=-9/2$ is produced by a matter domination scale factor, that is, $a(t)\sim t^{2/3}$. In principle, there might be other scale factors which may produce a constant parameter $m$, however the physical significance of each case is determined by the behavior of the equation of state parameter $w_{eff}$ and of the final attractor of the theory. The effective equation of state (EoS) can be written in terms of the variables $x_i$, $i=1,2,3$. The general form of the EoS for an $f(R)$ gravity has the following for \cite{reviews1,reviews2,reviews3,reviews4,reviews5},
\begin{equation}\label{weffoneeqn}
w_{eff}=-1-\frac{2\dot{H}}{3H^2}\, ,
\end{equation}
and it can be expressed in terms of $x_3$ in the following way,
\begin{equation}\label{eos1}
w_{eff}=-\frac{1}{3} (2 x_3-1)\, .
\end{equation}
Eventually, the form of the total EoS actually determines the physical significance of the cosmological evolution corresponding to some fixed constant $m$. For example, as was shown in Ref. \cite{Odintsov:2017tbc}, in the de Sitter or the quasi-de Sitter case which corresponds to $m=0$, after the final attractors are found for the dynamical system, the EoS is equal to $w_{eff}=-1$, and for the case $m=-9/2$ one has $w_{eff}=0$.

Coming back to the description of the dynamical system, the variables $x_i$, $i=1,2,3$ satisfy the Friedmann constraint,
\begin{equation}\label{friedmanconstraint1}
x_1+x_2+x_3=1\, ,
\end{equation}
as it can be seen from Eq. (\ref{JGRG15}), by reexpressing $R=12H^2+6\dot{H}$.

The structure of the phase space for the $m=0$ is particularly interesting as was shown in \cite{Odintsov:2017tbc}, and here we briefly present the results of the analysis of Ref. \cite{Odintsov:2017tbc}. For general values of $m$, the fixed points of the dynamical system are,
\begin{align}\label{fixedpointsgeneral}
& \phi_*^1=(\frac{1}{4} \left(-\sqrt{-2m}-\sqrt{-2 m+20 \sqrt{2}
\sqrt{-m}+4}-2\right),\frac{1}{4} \left(3 \sqrt{2}
\sqrt{-m}+\sqrt{-2 m+20 \sqrt{-2m}+4}-2\right),\frac{1}{2}
\left(4-\sqrt{-2m}\right)),
 \\ \notag & \phi_*^2=(\frac{1}{4} \left(-\sqrt{-2m}+\sqrt{-2 m+20 \sqrt{-2m}+4}-2\right),\frac{1}{4} \left(3 \sqrt{-2m}-\sqrt{-2 m+20 \sqrt{-2m}+4}-2\right),\frac{1}{2} \left(4-\sqrt{-2m}\right)),
 \\ \notag & \phi_*^3=(\frac{1}{4} \left(\sqrt{-2m}-\sqrt{-2 m-20 \sqrt{-2m}+4}-2\right),\frac{1}{4} \left(-3 \sqrt{-2m}+\sqrt{-2 m-20 \sqrt{-2m}+4}-2\right),\frac{1}{2} \left(\sqrt{-2m}+4\right))
 \\ \notag & \phi_*^4=(\frac{1}{4} \left(\sqrt{-2m}+\sqrt{-2 m-20 \sqrt{-2m}+4}-2\right),\frac{1}{4} \left(-\sqrt{2} \sqrt{-m-10 \sqrt{-2m}+2}-3 \sqrt{-2m}-2\right),\frac{1}{2} \left(\sqrt{2}
 \sqrt{-m}+4\right))\, ,
\end{align}
so in the case $m\simeq 0$, the fixed points are the following,
\begin{equation}\label{fixedpointdesitter}
\phi_*^1=(x_1,x_2,x_3)=(-1,0,2),\,\,\,\phi_*^2=(x_1,x_2,x_3)=(0,-1,2)\, .
\end{equation}
It is easy to show that none of the above fixed points are hyperbolic, so the stability analysis must be performed numerically, as was done in Ref. \cite{Odintsov:2017tbc}. Let us point out that a hyperbolic fixed point has a Jacobian linearized matrix which has eigenvalues corresponding to the fixed point that have non-zero real part. More formally, a hyperbolic fixed point
does not have any center manifolds. We shall use the results of \cite{Odintsov:2017tbc}, where it was shown that in the $m\simeq 0$ case, the first fixed point $\phi_*^1$ is stable, while the second one $\phi_*^2$ is unstable. These two results shall play a crucial role in a mathematically non-trivial problem that will occur at the next section, however, since our results will be analytical, the resulting picture is both physically and mathematically clear and appealing. The equilibrium that will play a crucial role in one of the following sections, is the stable one, namely $\phi_*^1=(x_1,x_2,x_3)=(-1,0,2)$, which as we shall see will be related to the Big-Rip singularity. For this fixed point, the variable $x_3$ tends to the value $x_3=2$, as the fixed point is approached by the dynamical system, and by substituting $x_3=2$ in the EoS Eq. (\ref{eos1}), we obtain $w_{eff}=-1$. So the fixed point $\phi_*^1$ describes an accelerating physical evolution, in cosmological terms.

\section{Finite-time Singularities in the Dynamical System of Vacuum $f(R)$ Cosmology}

We shall be interested in understanding the structure of the vacuum $f(R)$ gravity phase space near finite-time singularities, so we assume for simplicity that the Hubble rate near the singularity has the following form,
\begin{equation}\label{hubblerate}
H(t)\simeq f_0(t_s-t)^{-\alpha}\, ,
\end{equation}
where $t_s$ is the time instance that the future singularity occurs, $\alpha$ is some real parameter, the value of which will determine the type of singularity that may occur, and also $f_0$ is a free parameter with dimensions sec$^{^{\alpha-1}}$. Also we assume that $t_s>t$, since the singularity is a future singularity. The classification of finite-time singularities was systematically performed in Ref. \cite{Nojiri:2005sx} and according to the classification of Ref. \cite{Nojiri:2005sx}, three physical quantities determine the type of singularity, namely the scale factor of the Universe, the total effective energy density $\rho_{\mathrm{eff}}$ and the total effective pressure $p_{\mathrm{eff}}$. The classification of finite-time singularities is as follows,
\begin{itemize}
\item Type I (``Big Rip'') : This is a characteristic crushing type singularity, for which as $t \to t_s$, the scale factor, the total effective pressure and the total effective energy density diverge strongly, that is, $a \to \infty$, $\rho_\mathrm{eff} \to \infty$, and $\left|p_\mathrm{eff}\right| \to \infty$. For extensive works on this type of singularity, the reader is referred in Refs. \cite{bigrip}.
\item Type II (``sudden''): This type of singularity is more mild than the Big Rip scenario, and it is also known as a pressure singularity, firstly studied in Refs. \cite{barrowsudden}, and later developed in \cite{barrowsudden1}, see also \cite{Balcerzak:2012ae,Marosek:2018huv}. In this case, only the total effective pressure diverges as $t \to t_s$, and the total effective energy density and the scale factor remain finite, that is, $a \to a_s$, $\rho_\mathrm{eff} \to \rho_s$, $\left|p_\mathrm{eff}\right| \to \infty$.
\item Type III : In this type of singularity, both the total effective pressure and the total effective energy density diverge as $t \to t_s$, but the scale factor remains finite, that is, $a \to a_s$, $\rho_\mathrm{eff} \to \infty$, $\left|p_\mathrm{eff}\right| \to \infty$.
\item Type IV : This type of singularity is the mildest from a phenomenological point of view and recent studies were devoted on this sort of singularity \cite{Nojiri:2005sr,Nojiri:2005sx,Nojiri:2004pf,Barrow:2015ora,Nojiri:2015fra,Nojiri:2015fia,Odintsov:2015zza,Oikonomou:2015qha,Kleidis:2017ftt,Oikonomou:2015qfh}. In this case, all the aforementioned physical quantities remain finite as $t\to t_s$, that is $a \to a_s$, $\rho_\mathrm{eff} \to \rho_s$,
$\left|p_\mathrm{eff}\right| \to p_s$, and only the higher derivatives of the Hubble rate $\frac{\mathrm{d}^nH}{\mathrm{d}t^n}$, $n\geq 2$ diverge. These singularities have interesting implications for the phenomenological dynamics of the inflationary era, since the Universe may smoothly pass through these without any catastrophic implications on the physical quantities that can be defined on the three-dimensional spacelike  hypersurface defined by the time $t=t_s$, if the singularity occurs during the inflationary era \cite{Odintsov:2015gba}. As was shown in \cite{Odintsov:2015gba}, the graceful exit from the inflationary era may be triggered by this type of soft singularity, see Refs. \cite{Odintsov:2015gba} for details on these issues.
\end{itemize}
According to the classification of singularities given above, the singularities that may occur for the various values of $\alpha$ for the Hubble rate in Eq. (\ref{hubblerate}), are classified as follows,
\begin{itemize}
\item $\alpha > 1$ corresponds to the Type I singularity.
\item $0<\alpha<1$ corresponds to Type III singularity.
\item $-1<\alpha<0$ corresponds to Type II singularity.
\item $\alpha < -1$ corresponds to Type IV singularity.
\end{itemize}
As we will show, the values of the parameter $\alpha$ will crucially affect the functional form of the $f(R)$ gravity near the singularity.

Let us see how the dynamical system (\ref{dynamicalsystemmain}) becomes if we choose the Hubble rate as in Eq. (\ref{hubblerate}). The study of the resulting dynamical system will reveal the dynamics of the cosmological system near each type of finite-time singularity. First we express the cosmic time as function of the $e$-foldings number, by using the definition of the $e$-foldings number $N$,
\begin{equation}\label{efoldingsdefinition}
N=\int^tH(t)\mathrm{d}t\, ,
\end{equation}
and by substituting the Hubble rate (\ref{hubblerate}) in the above equation and solving with respect to the cosmic time, we obtain,
\begin{equation}\label{tsasfunctionofefoldings}
t_s-t=\left(\frac{(\alpha -1) (N-N_c)}{f_0}\right)^{\frac{1}{1-\alpha }}\, ,
\end{equation}
where $N_c$ is an integration constants which depends on the initial conditions. The above relation is very important for the analysis that follows in the next subsections, since depending on the values of the parameter $\alpha$, this relation will determine the values of $N$ as the corresponding singularity is approached. For the moment though we leave the value of $\alpha$ unspecified. By calculating the value of the parameter $m$ given in Eq. (\ref{parameterm}), for the Hubble rate (\ref{hubblerate}), and by using Eq. (\ref{tsasfunctionofefoldings}), we obtain the parameter $m$ as function of the $e$-foldings number $N$, which is,
\begin{equation}\label{parametermasfunctionofN}
m=-\frac{\alpha  (\alpha +1)}{(\alpha -1)^2 (N_c-N)^2}\, .
\end{equation}
The above relation for $m$ holds true for all the cosmological finite-time singularity types, regardless the value that $\alpha$ takes, however the difference will be that for each singularity type, $N$ takes different values, that is, as $t\to t_s$, $N$ takes different values for the various singularity types. By substituting the parameter $m$ in the dynamical system (\ref{dynamicalsystemmain}), the latter becomes,
\begin{align}\label{dynamicalsystemnonautonomousgeneral}
& \frac{\mathrm{d}x_1}{\mathrm{d}N}=-4+3x_1+2x_3-x_1x_3+x_1^2\, ,
\\ \notag &
\frac{\mathrm{d}x_2}{\mathrm{d}N}=-\frac{\alpha  (\alpha +1)}{(\alpha -1)^2 (N_c-N)^2}+8-4x_3+x_2x_1-2x_2x_3+4x_2 \, ,\\
\notag & \frac{\mathrm{d}x_3}{\mathrm{d}N}=\frac{\alpha  (\alpha +1)}{(\alpha -1)^2 (N_c-N)^2}-8+8x_3-2x_3^2 \, .
\end{align}
The dynamical system of Eq. (\ref{dynamicalsystemnonautonomousgeneral}) is non-autonomous, so most of the theorems that apply for autonomous dynamical systems are rendered inapplicable in the case at hand. In principle it is quite difficult to tackle the non-autonomous system, however the above dynamical system can be integrated analytically in some limiting cases. The behavior of the solutions then, will reveal the structure of the phase space of vacuum $f(R)$ gravity, near finite-time singularities.

The focus in this section is two-fold, firstly we shall be interested in the behavior of the vacuum $f(R)$ dynamical system near the singularity, and we shall analyze each singularity type separately, and secondly we shall investigate what is the functional form of the $f(R)$ which may generate each type of singularity, directly from the data obtained from the phase space itself. As we shall see, the dynamical system can be integrated analytically for the Hubble rate (\ref{hubblerate}), for some limiting values of $\alpha$, so this offers us great flexibility for obtaining the structure of the phase space near the singularity. Also, the resulting behavior near the singularity for the Big Rip case will enable us to obtain an interesting behavior for the non-autonomous system asymptotically for large $N$. In the next subsections we analyze the various types of singularities separately, with the cases that can be analyzed analytically though, being the Big Rip and the Type IV singularity.

\subsection{The Big-Rip Case}

We start off our analysis with the Big Rip singularity case, which is the most severe type of singularity, from a phenomenological point of view, since it is a crushing type singularity following the definitions of Hawking and Penrose \cite{hawkingpenrose}. Our choice to study the Big Rip case separately from the other singularities is not accidental, since as we now demonstrate, in this case the dynamical system has a mathematically interesting property, that is, it is rendered asymptotically autonomous near the singularity.

In the Big Rip case, $\alpha$ in the Hubble rate (\ref{hubblerate}) must take values $\alpha>1$, so as $t\to t_s$, from Eq. (\ref{tsasfunctionofefoldings}) it follows that this corresponds to $N\to \infty$. Hence, the Big Rip singularity is approached as $N\to \infty$. For $N\to \infty$, the parameter $m$ in Eq. (\ref{parametermasfunctionofN}) tends to $m\to 0$, hence the dynamical system (\ref{dynamicalsystemnonautonomousgeneral}) near the Big Rip singularity is rendered asymptotically autonomous. This is an exceptional case and it occurs only for the Big Rip singularity. Actually, the fact that the $f(R)$ gravity dynamical system is asymptotically autonomous has some interesting mathematical implications, as we shall briefly discuss at the end of this subsection.

\begin{figure}[h]
\centering
\includegraphics[width=16pc]{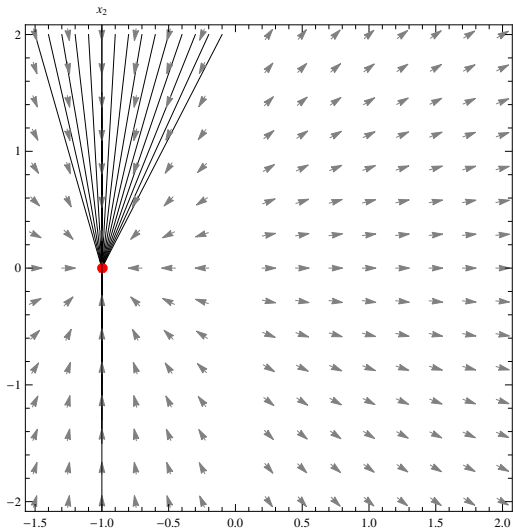}
\caption{{\it{The vector flow in the $x_1-x_2$ plane for the dynamical system that is obtained for $x_3=2$. The red dot indicates the point $(x_1,x_2)=(-1,0)$.}}} \label{plot1newa}
\end{figure}

\begin{figure}[h]
\centering
\includegraphics[width=16pc]{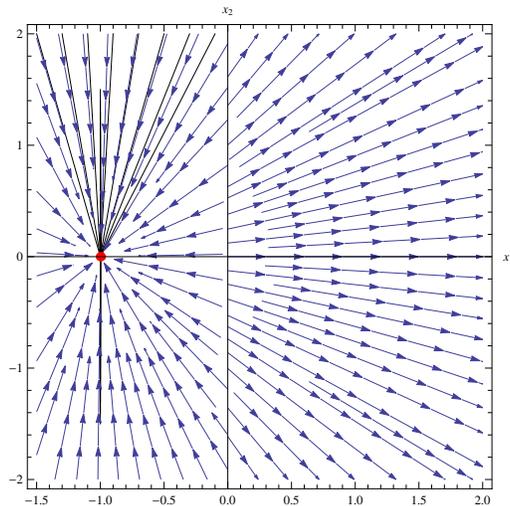}
\caption{{\it{The trajectories in the $x_1-x_2$ plane for $x_3=2$, for various initial conditions near $(x_1,x_2)=(-1,0)$ (upper right). The red dot indicates the point $(x_1,x_2)=(-1,0)$.}}} \label{plot1newb}
\end{figure}

\begin{figure}[h]
\centering
\includegraphics[width=16pc]{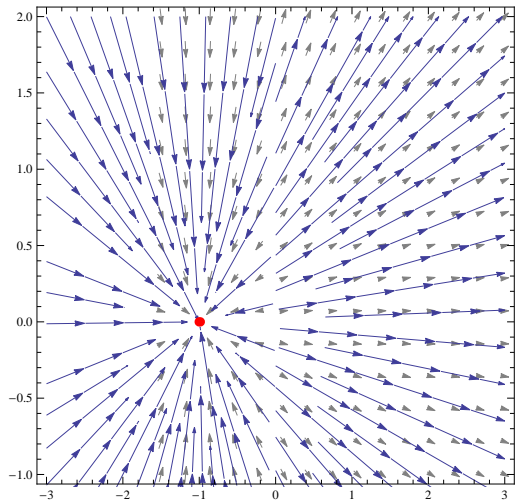}
\caption{{\it{The flows in the $x_1-x_2$ plane (bottom plot). The red dot indicates the point $(x_1,x_2)=(-1,0)$.}}} \label{plot1newc}
\end{figure}

\begin{figure}[h]
\centering
\includegraphics[width=20pc]{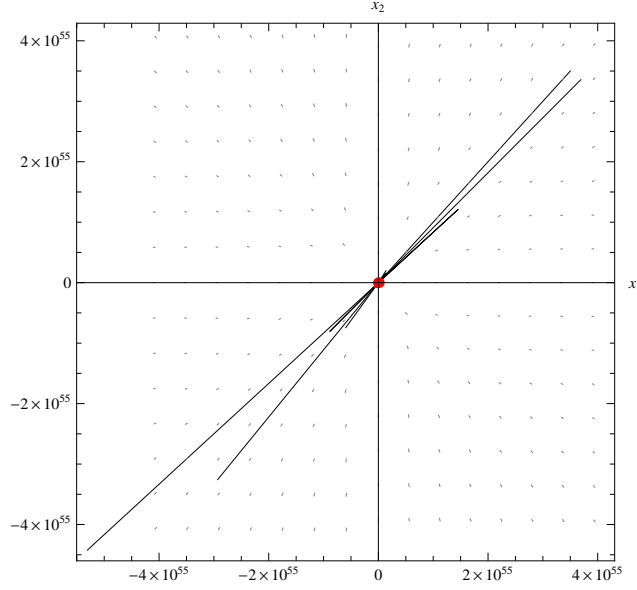}
\caption{{\it{The vector flow and trajectories in the $x_1-x_2$ plane for large initial conditions.}}} \label{plot2}
\end{figure}

A particularly interesting scenario occurs if $\alpha$ is chosen to take large values, that is $\alpha\gg 1$. In this case, the dynamical system can be integrated analytically, and the parameter $m$ becomes approximately $m\simeq -\frac{1}{(N-N_c)^2}$. By integrating the dynamical system (\ref{dynamicalsystemnonautonomousgeneral}), we obtain the following general solutions for the variables $x_1(N)$, $x_2(N)$ and $x_3(N)$,
\begin{align}\label{generalsolutionsforx1x2x3}
& x_1(N)= \frac{-\frac{3 \sqrt{\pi } N_c e^N \mathrm{Erf}\left(\sqrt{N-N_c}\right)}{\sqrt{N-N_c}}+2 e^{N_c} N-2 e^{N_c} N_c+2 e^{N_c}}{e^{N_c} (2 N_c-2 N+1)-2 \sqrt{\pi } e^N (N-N_c)^{3/2} \mathrm{Erf}\left(\sqrt{N-N_c}\right)}
\\ \notag & +\frac{\frac{2 \sqrt{\pi } N_c^2 e^N \mathrm{Erf}\left(\sqrt{N-N_c}\right)}{\sqrt{N-N_c}}+\frac{2 \sqrt{\pi } e^N N \mathrm{Erf}\left(\sqrt{N-N_c}\right)}{\sqrt{N-N_c}}}{e^{N_c} (2 N_c-2 N+1)-2 \sqrt{\pi } e^N (N-N_c)^{3/2} \mathrm{Erf}\left(\sqrt{N-N_c}\right)}
\\ \notag &
-\frac{\frac{2 \sqrt{\pi } e^N N^2 \mathrm{Erf}\left(\sqrt{N-N_c}\right)}{\sqrt{N-N_c}}+\frac{4 \sqrt{\pi } N_c e^N N \mathrm{Erf}\left(\sqrt{N-N_c}\right)}{\sqrt{N-N_c}}}{e^{N_c} (2 N_c-2 N+1)-2 \sqrt{\pi } e^N (N-N_c)^{3/2} \mathrm{Erf}\left(\sqrt{N-N_c}\right)}
\\ \notag & +\frac{\sqrt{\pi } e^N N \mathrm{Erf}\left(\sqrt{N-N_c}\right)}{\sqrt{N-N_c} \left(e^{N_c} (2 N_c-2 N+1)-2 \sqrt{\pi } e^N (N-N_c)^{3/2} \mathrm{Erf}\left(\sqrt{N-N_c}\right)\right)}
\\ \notag &
x_2(N)= \frac{\mathcal{C}_2 e^{N_c-N} (N_c-N)}{2 \sqrt{\pi } (N-N_c)^{3/2} \mathrm{Erf}\left(\sqrt{N-N_c}\right)+e^{N_c-N} (2 N_c-2 N+1)}\\ \notag &
-\frac{e^{-N} (N_c-N) \left(\frac{e^{N_c} (8 N_c-8 N+1)}{2 (N_c-N)^2}-\frac{4 \sqrt{\pi } e^N \mathrm{Erf}\left(\sqrt{N-N_c}\right)}{\sqrt{N-N_c}}\right)}{2 \sqrt{\pi } (N-N_c)^{3/2} \mathrm{Erf}\left(\sqrt{N-N_c}\right)+e^{N_c-N} (2 N_c-2 N+1)}
\\ \notag &
x_3(N)=\frac{1}{2 N_c-2 N}+2\, ,
\end{align}
where $\mathrm{Erf}(x)$ is the error function, and also $\mathcal{C}_2$ is an integration constant. Actually, the dynamical system can be integrated analytically, only if we first integrate the equation for the variable $x_3$, and then we solve the differential equations for $x_1$ and $x_2$. Also, asymptotically, the integration constant $\mathcal{C}_2$ will play no essential role in the physical behavior of the parameter $x_1$. We can find easily the asymptotic forms of the variables $x_1(N)$, $x_2(N)$ and $x_3(N)$ as $N\to \infty$, and these are the following,
\begin{align}\label{asymptoticformsofthevariables}
& x_1(N)\simeq -\frac{(N_c-N)^2}{N^2}\simeq -1,\\ \notag &
x_2(N)\simeq -\frac{2}{N}\simeq 0, \\ \notag &
x_3(N)\simeq 2\, .
\end{align}
Let us discuss in some detail the behavior of the dynamical system corresponding to Eq. (\ref{asymptoticformsofthevariables}). From Eq. (\ref{asymptoticformsofthevariables}), it is obvious that as $N\to \infty$, so effectively as the Big Rip singularity is approached, the variables $x_1(N)$, $x_2(N)$ and $x_3(N)$ take the values $(x_1,x_2,x_3)=(-1,0,2)$. By looking Eq. (\ref{fixedpointdesitter}), which describes the fixed points of the autonomous dynamical system (\ref{dynamicalsystemmain}) for $m=0$, it is obvious that the non-autonomous dynamical system (\ref{dynamicalsystemnonautonomousgeneral}) has asymptotically (as $N\to \infty$) the same fixed point with the autonomous dynamical system (\ref{dynamicalsystemmain}). Although this would seem natural to occur, this is not a trivial mathematical feature. Indeed, there are mathematically stringent conditions in order for a fixed point of a non-autonomous dynamical system which is also an asymptotically autonomous dynamical system,  to be the fixed point of the asymptotically autonomous system. In simple words, this means that the fixed points of the dynamical system (\ref{dynamicalsystemnonautonomousgeneral}) are not necessarily the same with the ones corresponding to the dynamical system  (\ref{dynamicalsystemmain}) for $m=0$. However in our case, our analytic treatment proved this feature directly, so this has important implications for the structure of the non-autonomous system, which we shall analyze in a future work, focused on this issue. The important feature for our analysis is that the non-autonomous system of Eq. (\ref{dynamicalsystemnonautonomousgeneral}), near the Big Rip singularity, tends to the attractor $\phi_*^1=(x_1,x_2,x_3)=(-1,0,2)$, which is a stable fixed point of the asymptotically autonomous system (\ref{dynamicalsystemmain}) for $m\simeq 0$. Hence, due to  the fact that asymptotically we have $x_3\simeq 2$, the EoS of the vacuum $f(R)$ gravity cosmological system (\ref{eos1}) is approximately $w_{eff}\simeq -1$. In effect, the phase space structure indicates that the Big Rip singularity for vacuum $f(R)$ gravity is inherently related to an accelerating period with $w_{eff}\simeq -1$. It is worth to further study the behavior of the phase space, so in Fig. \ref{plot1newa}, we present the vector flow and trajectories in the $x_1-x_2$ plane for the dynamical system that is obtained for $x_3=2$, for various initial conditions near the point $(x_1,x_2)=(-1,0)$. Accordingly, in Fig. \ref{plot1newb} we present the stream flow and the trajectories in the $x_1-x_2$ plane for $x_3=2$, and in Fig. \ref{plot1newc} we present the flows in the $x_1-x_2$ plane. In all cases, the red dot indicates the point $(x_1,x_2)=(-1,0)$. As it can be seen, the trajectories of dynamical system tend to the point $(x_1,x_2)=(-1,0)$, exactly as we described analytically before.
In Fig. \ref{plot2} we plot some trajectories which correspond to initial conditions which are not near  the point $(x_1,x_2)=(-1,0)$. As it can be seen, the trajectories do not end up to the point $(x_1,x_2)=(-1,0)$. This feature shows that there exist trajectories corresponding to some initial conditions not in the close neighborhood of $(x_1,x_2)=(-1,0)$, which do not lead to the point $(x_1,x_2)=(-1,0)$. This feature cannot directly be connected with the stability of the fixed point $(x_1,x_2)=(-1,0)$ of the asymptotically autonomous system, only the close neighborhood of the point $(x_1,x_2)=(-1,0)$ captures the dynamics of the dynamical system near the Big Rip singularity.
Having at hand the analytic form of the variables $x_i$, $i=1,2,3$, at large $N$, which is $(x_1,x_2,x_3)=(-1,0,2)$, we can find the functional form of the $f(R)$ gravity near the Big Rip singularity. Taking into account that $x_1\simeq -1$, we obtain the following differential equation,
\begin{equation}\label{diffeqx1}
-\frac{\dot{F}}{FH}=-1\, ,
\end{equation}
so by using Eq. (\ref{hubblerate}) and also due to the fact that near the Big Rip singularity the Ricci scalar is at leading order equal to,
\begin{equation}\label{leadingorderricciscalar}
R(t)\simeq 12 f_0^2 (t_s-t)^{-2 \alpha }\, ,
\end{equation}
we obtain the following solution for $F(R)$,
\begin{equation}\label{frbigripcase}
F(R)\simeq \exp \left(\gamma  R^{\frac{\alpha -1}{2 \alpha }}\right)+\Lambda_I\, ,
\end{equation}
where $\Lambda_I$ is an integration constant and also $\gamma$ is equal to,
\begin{equation}\label{gamma}
\gamma=\frac{f_0}{(\alpha-1)(12f_0^2)^{\frac{\alpha-1}{2\alpha}}}\, .
\end{equation}
By integrating Eq. (\ref{frbigripcase}) with respect to the Ricci scalar, we obtain the functional form of the $f(R)$ gravity near the Big Rip singularity, which is,
\begin{equation}\label{frfinalbigrip1}
f(R)\simeq \Lambda_{I}\,R+\frac{2 \alpha  \gamma ^{-\frac{2 \alpha }{\alpha -1}} \Gamma \left(\frac{2 \alpha }{\alpha -1},-R^{\frac{\alpha -1}{2 \alpha }} \gamma \right)}{\alpha -1}+\Lambda_{II}\, ,
\end{equation}
where $\Lambda_{II}$ is an integration constant. Due to the fact that as $t\to t_s$, the Ricci scalar tends to $R\to \infty$, and also since $\alpha>1$, the $f(R)$ gravity (\ref{frfinalbigrip1}) at leading order is equal to,
\begin{equation}\label{frfinalbigrip2}
f(R)\simeq \Lambda_{I}\,R-\frac{\left(2 \alpha  \gamma ^{\frac{\alpha +1}{\alpha -1}-\frac{2 \alpha }{\alpha -1}}\right) R^{\frac{\alpha +1}{2 \alpha }} e^{\gamma  R^{\frac{\alpha -1}{2 \alpha }}}}{\alpha -1}+\Lambda_{II}\, ,
\end{equation}
where we required that for consistency, $\alpha$ has the form $\alpha=2n/(2m+1)$, with $n$ and $m$ integers. We need to stress that the functional behavior of the $f(R)$ gravity appearing in Eq. (\ref{frfinalbigrip2}) is very similar to the one found in Refs. \cite{Bamba:2008ut,Nojiri:2008fk}. Note that Eqs. (\ref{frfinalbigrip1}) and (\ref{frfinalbigrip2}) are valid in the regime $\alpha \gg 1$ and not in general. Also in this regime, from Eq. (\ref{frfinalbigrip2}) the approximate form of the $f(R)$ gravity is,
\begin{equation}\label{mtouc}
f(R)\simeq \Lambda_{I}\,R-\frac{\left(2 \alpha  \gamma ^{-1}\right) R^{\frac{1}{2}} e^{\gamma  R^{\frac{1}{2}}}}{\alpha}+\Lambda_{II}\, ,
\end{equation}
so the resulting $f(R)$ gravity is similar to the Einstein-Hilbert case, but not identical, due to the existence of the term $\sim R^{\frac{1}{2}} e^{\gamma  R^{\frac{1}{2}}}$. Near the Big Rip, due to the fact that the curvature increases, the term containing the exponential $\sim e^{\gamma  R^{\frac{1}{2}}}$ will dominate the $f(R)$ gravity function (\ref{mtouc}) and hence it controls the evolution.

In the general case of singularities, the analytical treatment of the dynamical evolution might be difficult to tackle and may crucially  depend on the initial conditions, however for the simple form of the Hubble rate (\ref{hubblerate}), the calculations are simplified and an approximate form of the $f(R)$ near the singularity can be found, as we showed in detail in this subsection.

Before closing this subsection let us recapitulate the most important findings we obtained: firstly, as the Big Rip singularity is reached, for $N\to \infty$, the vacuum $f(R)$ gravity dynamical systems becomes asymptotically autonomous. Secondly, the analytic study of the non-autonomous dynamical system, for $\alpha\gg 1$, revealed that the non-autonomous dynamical system tends asymptotically to the stable fixed point of the asymptotically autonomous dynamical system. This is a particularly interesting feature from a mathematical point of view, which we shall analyze in a future focused work. Finally, we proved that the asymptotic behavior of the $f(R)$ gravity dynamical system tends to an accelerating attractor point, so essentially the Big Rip singularity is inherently related with the acceleration of the Universe.

What we did not address in this subsection, is the effect of an $R^2$ term in the functional form of the $f(R)$ gravity. The addition of an $R^2$ term in a singularity generating $f(R)$ gravity may crucially affect the singularity development in the theory, since it is known that these terms remove the singularities \cite{Bamba:2008ut,Nojiri:2008fk,Capozziello:2009hc,Nojiri:2009pf,Elizalde:2010ts}. Also the addition of an $R^2$ term in the gravitational action can also describe the large curvature era during inflation, so the possible unification of the dark energy era with the inflationary era is possible, see \cite{Bamba:2008ut,Nojiri:2008fk,Capozziello:2009hc,Nojiri:2009pf,Elizalde:2010ts}. However, after some thorough analysis at the dynamical system level, the answer to this question is not easy to find, due to the fact that the resulting system is very difficult to solve analytically, or even study numerically, due to the stiffness of the resulting differential equations. Hence, some analytical insight is needed to address formally this issue, however this task exceeds by far the purposes of this article, so we postpone this task to a future work.

Also, a Big Rip singularity in the $f(R)$ frame may be related to phantom crossing or a Type IV singularity in the Einstein frame \cite{Bamba:2008hq}.

\subsection{The Type III, Type II and Type IV Cases}

Let us now focus on the Type III, Type II and Type IV cases, for which as $t\to t_s$, from Eq. (\ref{tsasfunctionofefoldings}) it follows that $N\to N_c$. Since in the above cases, the cosmic time can also take values $t>t_s$, we additionally require that $\alpha$ has the form,
\begin{equation}\label{alpjarequirements}
\alpha=\frac{2m}{2n+1}\, ,
\end{equation}
where $n$ and $m$ are positive integers.

We start off with the Type III singularity, and due to the fact that the singularity is approached as $N\to N_c$, the parameter $m$ appearing in Eq. (\ref{parametermasfunctionofN}) diverges. Unfortunately it is very difficult to obtain an analytic solution for the dynamical system (\ref{dynamicalsystemmain}), for the Type III and Type II cases. However, for the Type IV singularity, analytic results can be obtained if $\alpha\ll -1$. So we focus on the Type IV case, and for $\alpha\ll -1$, the parameter $m$ appearing in Eq. (\ref{parameterm}) becomes $m\simeq -\frac{1}{(N-N_c)^2}$. In this case, the dynamical system (\ref{dynamicalsystemnonautonomousgeneral}) has the same solutions $x_1(N)$, $x_2(N)$ and $x_3(N)$ as the ones appearing in Eq. (\ref{generalsolutionsforx1x2x3}), for general $N$, but in this case near the Type IV singularity at $N=N_c$, the parameters $x_1(N)$, $x_2(N)$ and $x_3(N)$ take the following form,
\begin{align}\label{asymptoticformsofthevariables1}
& x_1(N)\simeq \left(\frac{4 N_c}{3}+12\right) (N-N_c)+2,\\ \notag &
x_2(N)\simeq -\frac{1}{2(N-N_c)}-3-(\mathcal{C}_2-12)(N-N_c), \\ \notag &
x_3(N)= \frac{1}{2 N_c-2 N}+2\, ,
\end{align}
and in effect, as $N\to N_c$, the dynamical system reaches the point $(x_1,x_2,x_3)=(2,-\infty,\infty)$. Notice that although the dynamical variables $x_2$ and $x_3$ diverge, the singularities actually cancel, and hence the Friedmann constraint (\ref{friedmanconstraint1}) is satisfied.

In the case at hand, we can also find the functional form of the $f(R)$ gravity near the Type IV singularity, by solving the differential equation $x_1=2$, and by following the procedure of the previous subsection, and also using the fact that as $N\to N_c$, the dominant part of the scalar curvature is again given by,
\begin{equation}\label{dominanttypeivcurvature}
R\simeq 6f_0\alpha(t_s-t)^{-\alpha-1}\, .
\end{equation}
So the resulting $F(R)$ gravity is,
\begin{equation}\label{frtypeiiibig}
F(R)\simeq \Lambda_{III}+\exp \left(-\gamma_I  R^{-\frac{1-\alpha }{\alpha +1}}\right)\, ,
\end{equation}
where $\Lambda_{III}$ is an integration constant and $\gamma_I$ is,
\begin{equation}\label{gammaI}
\gamma_I=\frac{2f_0}{(1-\alpha)(6f_0 |\alpha|)^{-\frac{1-\alpha }{\alpha +1}}}\, .
\end{equation}
By integrating Eq. (\ref{frtypeiiibig}) with respect to the Ricci scalar, the resulting $f(R)$ gravity has the following form,
\begin{equation}\label{frfinaltypeiii}
f(R)\simeq \Lambda_{III}\,R+\frac{(\alpha +1) \gamma_I^{\frac{\alpha +1}{1-\alpha }} \Gamma \left(\frac{\alpha +1}{\alpha -1},R^{\frac{\alpha -1}{\alpha +1}} \gamma_I\right)}{1-\alpha }+\Lambda_{IV}\, ,
\end{equation}
where $\Lambda_{IV}$ is an integration constant. We can find the leading order behavior near the Type IV singularity, so by using the fact that as $N\to N_c$, the curvature goes to zero, due to the fact that $\alpha\ll -1$. So the leading order $f(R)$ gravity has the following form,
\begin{equation}\label{leadingorderfrtypeiii}
f(R)\simeq R+\Lambda_{III}\,R+\frac{(\alpha +1) \gamma_I^{\frac{\alpha +1}{1-\alpha }} \Gamma \left(\frac{\alpha +1}{\alpha -1}\right)}{1-\alpha }-\frac{(\alpha +1) \gamma_I R^{\frac{2 \alpha }{\alpha +1}}}{2 \alpha }+\Lambda_{IV}\, .
\end{equation}
Finally, by exploiting the fact that $\alpha\ll -1$, we finally get the leading order behavior of the $f(R)$ gravity near the extremely soft Type IV singularity, which is,
\begin{equation}\label{typeivfinalfr}
f(R)\simeq R+\Lambda_{III}\,R-\frac{1}{\gamma_I}-\frac{\gamma_I R^2}{2}+\Lambda_{IV}\, .
\end{equation}

\begin{figure}[h]
\centering
\includegraphics[width=18pc]{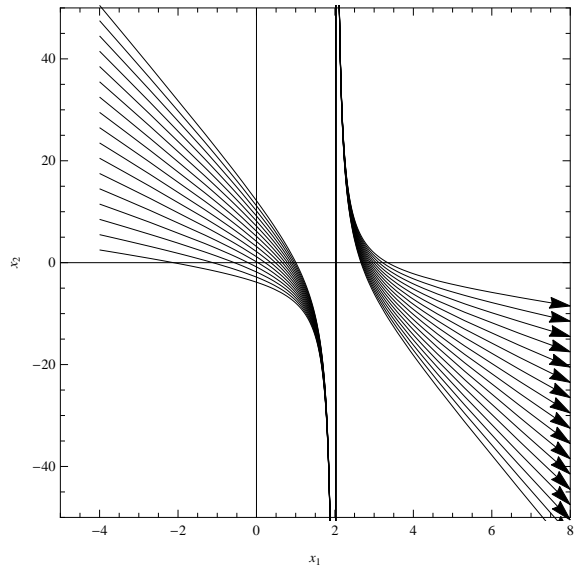}
\caption{{\it{The trajectories in the $x_1-x_2$ plane for different values of the free parameter $\mathcal{C}_2$ and for $N_c=12$. It can be seen that near the singularity at $N=N_c$, the variable $x_1$ tends to $x_1\to 2$, while $x_2$ tends to infinity.}}} \label{plot3newa}
\end{figure}

\begin{figure}[h]
\centering
\includegraphics[width=18pc]{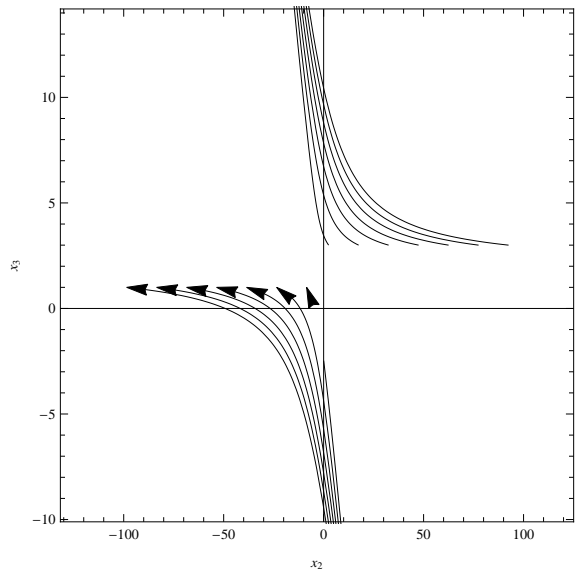}
\caption{{\it{The trajectories in the $x_2-x_3$ plane, for different values of the free parameter $\mathcal{C}_2$ and for $N_c=12$. It can be seen that near the singularity at $N=N_c$, the variables $x_2$ and $x_3$ tend to infinity.}}} \label{plot3newb}
\end{figure}

\begin{figure}[h]
\centering
\includegraphics[width=18pc]{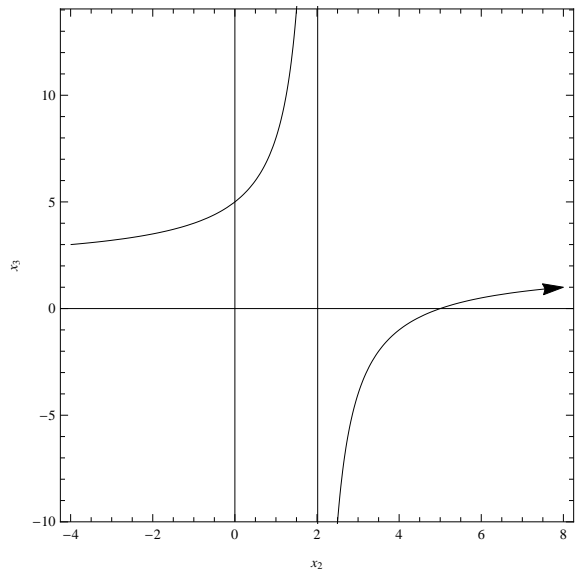}
\caption{{\it{The trajectories in the $x_1-x_3$ plane, for different values of the free parameter $\mathcal{C}_2$ and for $N_c=12$. It can be seen that near the singularity at $N=N_c$, the variables $x_2$ and $x_3$ tend to infinity.}}} \label{plot3newc}
\end{figure}

The resulting phase space of the dynamical system is particularly interesting near the Type IV singularity, since the variables $x_2$ and $x_3$ blow-up as $N\to N_c$, while $x_1\to 2$. This can be seen in the phase plots, so in Fig. \ref{plot3newa} we plot the trajectories in the $x_1-x_2$ plane, in Fig. \ref{plot3newb} the trajectories in the $x_2-x_3$ plane and in Fig. \ref{plot3newc} one trajectory in the $x_2-x_3$ plane, for different values of the free parameter $\mathcal{C}_2$ and for $N_c=12$ (the value of $N_c$ plays no important role in the qualitative behavior of the plots). As the singularity is approached, $x_1$ tends to $x_1\to 2$, while $x_2$ tends to infinity. This behavior is explained perfectly from the asymptotic behavior of the variables $x_1(N)$, $x_2(N)$ and $x_3(N)$ near the Type IV singularity, namely Eqs. (\ref{asymptoticformsofthevariables1}). Also in the $x_1-x_3$ plane, only one curve appears, and this is due to the fact that in the $x_2$ variable the free parameter $\mathcal{C}_2$  appears in the asymptotic expansion.

\subsection{The non-vacuum $f(R)$ Gravity Dynamical System Case: Problems and Difficulties}

In the previous subsections we studied the behavior of the vacuum $f(R)$ gravity near finite-time singularities, and in this subsection we discuss the case that matter is included in the $f(R)$ theory, and particularly non-relativistic matter and radiation. The $f(R)$ gravity gravitational equations for the FRW metric (\ref{frw}) become in this case,
\begin{align}
\label{JGRG15new} 0 =& -\frac{f(R)}{2} + 3\left(H^2 + \dot H\right)
F(R) - 18 \left( 4H^2 \dot H + H \ddot H\right) F'(R)+\kappa^2\rho_{matter}\, ,\\
\label{Cr4b} 0 =& \frac{f(R)}{2} - \left(\dot H + 3H^2\right)F(R) +
6 \left( 8H^2 \dot H + 4 {\dot H}^2 + 6 H \ddot H + \dddot H\right)
F'(R) + 36\left( 4H\dot H + \ddot H\right)^2 F'(R)+
\kappa^2p_{matter}\, ,
\end{align}
where $\rho_{matter}$ and $p_{matter}$ stand for the total effective energy density and total effective pressure respectively. We define the following variables,
\begin{equation}\label{variablesslowdownnew}
x_1=-\frac{\dot{F}(R)}{F(R)H},\,\,\,x_2=-\frac{f(R)}{6F(R)H^2},\,\,\,x_3=
\frac{R}{6H^2},\,\,\,x_4=\frac{\kappa^2\rho_r}{3FH^2},\,\,\,x_5=\frac{\kappa^2\rho_m}{3FH^2}\,
,
\end{equation}
with $\rho_r$ and $\rho_m$ being the radiation and matter energy
densities. By combining the gravitational equations (\ref{JGRG15new}) and the variables (\ref{variablesslowdownnew}), the dynamical system (\ref{dynamicalsystemmain}) becomes in this case,
\begin{align}\label{dynamicalsystemmain2}
&
\frac{\mathrm{d}x_1}{\mathrm{d}N}=-4+3x_1+2x_3-x_1x_3+x_1^2+3x_5+4x_4\,
,
\\ \notag &
\frac{\mathrm{d}x_2}{\mathrm{d}N}=8+m-4x_3+x_2x_1-2x_2x_3+4x_2 \, ,\\
\notag & \frac{\mathrm{d}x_3}{\mathrm{d}N}=-8-m+8x_3-2x_3^2 \, ,\\
\notag & \frac{\mathrm{d}x_4}{\mathrm{d}N}=x_4x_1-2x_4x_3 \, ,\\
\notag & \frac{\mathrm{d}x_5}{\mathrm{d}N}=x_5+x_5x_1-2x_5x_3 \, ,
\end{align}
with the parameter $m$ being the same as in Eq. (\ref{parameterm}). For the Hubble rate (\ref{hubblerate}), the parameter $m$ becomes identical with the one appearing in Eq. (\ref{parametermasfunctionofN}), and the dynamical system is altered accordingly. However, the resulting dynamical system cannot be solved analytically, even for the limiting cases of the parameter $\alpha$ which we studied in the previous subsection. Also it is not possible to find the behavior of the $f(R)$ dynamical system near the singularities in an analytical but also in a numerical way. An exception is the variable $x_3$, which behaves exactly as in the autonomous dynamical system case. This issue of non predictability of the phase space near a finite-time singularity is worth analyzing in more detail. In the way we expressed the variables $x_i$, both in the vacuum and non-vacuum $f(R)$ gravity, it is not possible to predict the behavior of the variables near a finite-time singularity.
This is due to the fact that the variables do not affect directly the singularity structure, meaning that even if a variable blows at some finite-time, this does not necessarily implies that a finite-time singularity is approached in general. For example $x_1$ in the vacuum $f(R)$ gravity case for the Big Rip tends to $x_1\to -1$, and for the rest of finite-time singularities it tends to $x_1\to 2$, as  the singularity was approached. Hence, except possibly for the variables $x_4$ and $x_5$ in Eq. (\ref{variablesslowdownnew}) which contain explicitly the energy density, the rest of the variables cannot offer much information for the behavior of the dynamical system near a finite-time singularity. Therefore, the novelty of our approach and of our construction of the dynamical system is based on the fact that we can provide analytically the behavior of the $f(R)$ gravity near a finite-time singularity, although the dynamical system is non-autonomous near the singularity. However, if non-analytic results cannot be obtained, one has no information at all on how the dynamical system behaves near a singularity, unless the variables contain the energy density or the scale factor. In this case, the phase space could possibly reveal any singularity which is a Type I, or Type III, but not the pressure and Type IV singularities, since at the finite-time singularity the corresponding variables would blow-up. There is though a much more formal procedure based on dominant balances and asymptotic expansions of $\psi$-series near a finite-time singularity \cite{goriely,barrowcotsakis}. This method is applicable in strictly autonomous dynamical systems, and therefore is inapplicable in the $f(R)$ gravity and other modified gravities cases. In the next section we shall be interested in coupled multifluid cosmologies, which result to autonomous dynamical systems and also the variables contain enough information to determine the finite-time singularity structure of the cosmological system, at least for the Type II and Big Rip cases.

\section{Interacting Multifluid Cosmology and Finite-time Singularities}

As we already discussed in the introduction, the infrastructure of the dark sector is up to date unknown, apart from the fact that it dominates and controls the evolution of the Universe at present. In addition, since no dark matter particle has been ever observed,  and moreover the nature of the dark energy is still a mystery, the assumption that these mysterious driving forces of our Universe may be modelled by cosmological fluids, can be considered as a correct assumption, unless some counter-proof is found. Also in general, an interaction between these fluids might be possible, if not compelling, since the dark energy sector critically affects the $\Omega_m$ dark matter component of the Universe. Of course, the interaction should be carefully chosen, since at early times may affect the primordial matter density perturbations \cite{Eingorn:2015rma,Koshelev:2009nj}.

For the purposes of this paper we shall assume the presence of three fluids, two of which interact and describe the dark sector, namely the dark matter fluid with energy density $\rho_m$, the dark energy fluid with energy density fluid $\rho_d$ and the baryonic perfect fluid. An interaction between the dark sectors shall be assumed, and also in the dark energy  fluid we shall incorporate some bulk viscosity effects. Our aim is to construct an autonomous dynamical system from the cosmological equations that the fluids satisfy and we examine the possibility that finite-time singularities may arise. To this end, we shall use the findings of Ref. \cite{goriely}, which were later further developed in \cite{barrowcotsakis}, which relate the behavior of a polynomial autonomous dynamical system near a singularity, with the dominant balances of the dominant terms of the dynamical system. In order to render this article self-contained, we shall briefly present the general outline of the results of the theory developed in \cite{goriely}, later on in this section. Firstly, let us demonstrate how it is possible to obtain a polynomial autonomous dynamical system from the three cosmological fluids.

For the flat FRW metric of Eq. (\ref{frw}), the cosmological equations read,
\begin{equation}\label{flateinstein}
H^2=\frac{\kappa^2}{3}\rho_{tot}\, ,
\end{equation}
where $\rho_{tot}$ is the total energy density of the cosmological fluids, which is $\rho_{tot}=\rho_m+\rho_d+\rho_b$, and $\kappa^2=8\pi G$, where $G$ is Newton's gravitational constant. Also we assumed a physical units system for which $\hbar=c=1$. From Eq. (\ref{flateinstein}) we easily obtain,
\begin{equation}\label{derivativeofh}
\dot{H}=-\frac{\kappa^2}{3}\left( \rho_m+\rho_d+\rho_b+p_{tot}\right)\, ,
\end{equation}
where $p_{tot}$ is the total pressure, which is equal to the pressure of the dark energy fluid, which we denote $p_d$, since the dark matter and the baryonic fluids are pressureless. For the dark energy fluid, we shall assume the following generalized equation of state \cite{Nojiri:2005sr},
\begin{equation}\label{darkenergyeos}
p_d=-\rho_d-A\kappa^4\rho_d^2\, ,
\end{equation}
where $A$ is a real dimensionless variable. The energy-momentum conservation law for the three fluids leads to the following continuity equations,
\begin{align}\label{continutiyequations}
& \dot{\rho}_b+3H\rho_b=0\,  \\ \notag &
\dot{\rho}_m+3H\rho_m=Q\, \\ \notag &
\dot{\rho}_d+3H(\rho_d+p_d)=-Q\, ,
\end{align}
where $Q$ is the interaction term between the dark energy and dark matter, and $p_d$ is the dark matter pressure, which is given in Eq. (\ref{darkenergyeos}). The sign of the interaction term $Q$ actually determines the energy flow between the dark sectors, so if $Q>0$ the dark energy sector loses energy and gives it to the dark matter, and if $Q<0$ the converse procedure takes place. We shall choose $Q$ to have the following form,
\begin{equation}\label{qtermform}
Q=3H(c_1\rho_m+c_2\rho_d)\, ,
\end{equation}
which is phenomenologically motivated \cite{CalderaCabral:2008bx,Pavon:2005yx,Quartin:2008px,Sadjadi:2006qp,Zimdahl:2005bk}, where $c_1$, $c_2$ are real constants which must have the same sign in order to obtain a physically viable model.

The generalized EoS (\ref{darkenergyeos}) for a Universe filled only with the dark energy fluid leads unavoidably to finite-time singularities, and specifically to a Type III singularity. However for the three fluids case, it is rather difficult, if not completely impossible, to answer the question if finite-time singularities occur, at least by solving the resulting Einstein field equations. So one novel feature of our approach is that we will shall try to reveal the finite-time singularity structure of the three fluids case, by using dynamical systems arguments. Of course, the dominant balances study has limitations, but in principle we can have generalized results for the finite-time singularities of the dynamical system. But the question is, how the finite-time singularities of the dynamical system may actually indicate if the cosmological system has finite-time singularities. This can occur if the variables of the dynamical system are appropriately chosen. We shall discuss this issue in detail now, since this is very important from a physical point of view. The dynamical system for the cosmological system at hand can be constructed based on the equations (\ref{flateinstein}), (\ref{derivativeofh}) and (\ref{continutiyequations}), so we choose the dimensionless variables of the dynamical system as follows,
\begin{equation}\label{variablesofdynamicalsystem}
x_1=\frac{\kappa^2\rho_d}{3H^2},\,\,\,x_2=\frac{\kappa^2\rho_m}{3H^2},\,\,\,x_3=\frac{\kappa^2\rho_b}{3H^2},\,\,\,z=\kappa^2H^2\, .
\end{equation}
The dynamical system variables $x_i$, $i=1,2,3$ satisfy the Friedmann equation, which is now expressed as the following constraint,
\begin{equation}\label{friedmannconstraint}
x_1+x_2+x_3=1\, .
\end{equation}
Moreover, the total equation of state $w_{eff}=\frac{p_d}{\rho_{tot}}$ of the cosmological system, can also be expressed in terms of the dynamical system variables (\ref{variablesofdynamicalsystem}) as follows,
\begin{equation}\label{equationofstatetotal}
w_{eff}=-x_1-3Ax_1^2z\, .
\end{equation}
By using the cosmological equations (\ref{flateinstein}), (\ref{derivativeofh}) and (\ref{continutiyequations}) in conjunction with the dynamical variables (\ref{variablesofdynamicalsystem}), after some algebra we obtain the following dynamical system for general $Q$,
\begin{align}\label{dynamicalsystemmultifluid}
& \frac{\mathrm{d}x_1}{\mathrm{d}N}=-\frac{\kappa^2Q}{3H^3}+9Ax_1^2z^2+3x_1x_2+3x_1x_3-9Azx_1^3\, , \\ \notag &
\frac{\mathrm{d}x_2}{\mathrm{d}N}=\frac{\kappa^2Q}{3H^3}-3x_2+3x_2^2+3x_2x_3-9Ax_1^2x_2z\, , \\ \notag &
\frac{\mathrm{d}x_3}{\mathrm{d}N}=-3x_3+3x_3^2+3x_3x_2-9Ax_1^2x_3z\, \\ \notag &
\frac{\mathrm{d}z}{\mathrm{d}N}=-3x_2z-3x_3z+9Ax_1^2z^2\, ,
\end{align}
where we used the $e$-foldings number $N$ (\ref{efoldingsdefinition}), to be the dynamical variable of the cosmological system. By choosing the interaction $Q$-term as in Eq. (\ref{qtermform}), two additional terms are obtained for the dynamical system (\ref{dynamicalsystemmultifluid}), namely,
\begin{equation}\label{additionalterms}
\frac{\kappa^2Q}{3H^3}=3c_1x_2+3c_2x_1\, ,
\end{equation}
so the dynamical system (\ref{dynamicalsystemmultifluid}) has an additional linear contribution for the variables $x_1$ and $x_2$.

\subsection{Dominant-Balance Singularity Analysis of Autonomous Dynamical Systems}

%subsection%%%%%%%%%%%%%%%%%%%%%%%%%%%%%%%%%%%%%%%%55dominant balances analysis presentation in brief, kovaleskaya matrix, small example,where the singularity occurs

A concrete analysis which is applied to all polynomial type autonomous dynamical systems of any finite dimension, was performed in Ref. \cite{goriely}, where the authors provided sufficient conditions on the finite-time singularity occurrence of an autonomous dynamical system. We shall call this framework ``dominant balance analysis'' for brevity hereafter. Later on, the results of Ref. \cite{goriely} were applied to cosmological singularities in Ref. \cite{barrowcotsakis}. In this subsection we shall introduce some notation and terminology of the dominant balance theory, and also we shall introduce some fundamental quantities related to the theory, without getting into details.

Consider a $n-$dimensional dynamical system of the form,
\begin{equation}\label{dynamicalsystemdombalanceintro}
\dot{x}=f(x)\, ,
\end{equation}
where $x$ is some real vector $\epsilon$ $R^n$, and $f(x)=\left(f_1(x),f_2(x),...,f_n(x)\right )$. A finite-time singularity of this dynamical system is a moving singularity, which is related to the initial conditions of $x$. A moving singularity is some singularity of the form $(t-t_c)^{-p}$, where $t_c$ is simply an integration constant. For example the differential equation $\frac{\mathrm{d}y}{\mathrm{d}x}=\frac{1}{x^2y^2}$, has a solution $y=(\frac{1}{x}-c)^{-1}$, where $c$ is an integration constant. The solution has a singularity which depends on the initial conditions, that is at $\frac{1}{x}=c$, so this is called a moving singularity. The procedure in order to find if the autonomous dynamical system (\ref{dynamicalsystemdombalanceintro}) has finite-time singularities is based on the decomposition (truncation) of the function in possible dominant and subdominant parts, which we denote $\hat{f}(x)$ and $\breve{f}(x)$, so that the dominant part of the dynamical system becomes,
\begin{equation}\label{dominantdynamicalsystem}
\dot{x}=\hat{f}(x)\, .
\end{equation}
Notice that there are many ways to select the dominant part of the function $f(x)$, and also only one polynomial term is allowed for each entry of the vector $\hat{f}(x)$. Now we write each component $x_i$, $i=1,2,...,n$ of the vector $x$ as follows,
\begin{equation}\label{decompositionofxi}
x_1(\tau)=a_1\tau^{p_1},\,\,\,x_2(t)=a_2\tau^{p_2},\,\,\,....,x_n(t)=a_n\tau^{p_n}\, ,
\end{equation}
and we require that the full solution $x$ can be written in $\psi$-series in terms of $\tau=t-t_c$, where $t_c$ is the singularity time instance. Without getting into unnecessary details, we substitute the forms of the $x_i$'s appearing in Eq. (\ref{decompositionofxi}), in Eq. (\ref{dominantdynamicalsystem}), and for each different choice of $\hat{f}$, we equate the powers of the resulting polynomials. This will determine the parameters $p_i$, $i=1,2,...,n$, with the only constraint being that only fractional numbers or integers are allowed as solutions. We then can form the vector $\vec{p}=(p_1,p_2,...,p_n)$, which shall be important in the method we will use. Then, we can find the parameters $a_i$, $i=1,2,...,n$, which are uniquely determined by equating the coefficients of the resulting polynomials. In this way, the vector $\vec{a}=(a_1,a_2,a_3,....,a_n)$ is formed, for each choice of $\hat{f}$, which we shall call dominant balance. Only the non-zero dominant balances are allowed, and also note that the dominant balances may be complex numbers too. The vectors $\vec{a}$ and $\vec{p}$ form a balance $(\vec{a},\vec{p})$, this is why the method is called dominant balance method for detecting the possible finite-time singularities of an autonomous dynamical system.

The theorem developed by Goriely and Hyde in \cite{goriely}, indicates that if the dominant balance has complex entries, then the autonomous dynamical system (\ref{dynamicalsystemdombalanceintro}) has no finite-time singularities. If all the dominant balances are real, then finite-time singularities occur, however there is a stringent rule that determines if the singular solution is a general solution, that is, it is reached for all initial conditions, or it is a specific solution, which is reached for a small set of initial conditions. In this case, one computes the Kovalevskaya matrix $R$, which is defined as follows,
\begin{equation}\label{kovaleskaya}
R=\left(%
\begin{array}{ccccc}
  \frac{\partial \hat{f}_1}{\partial x_1} & \frac{\partial \hat{f}_1}{\partial x_2} & \frac{\partial \hat{f}_1}{\partial x_3} & ... & \frac{\partial \hat{f}_1}{\partial x_n} \\
  \frac{\partial \hat{f}_2}{\partial x_1} & \frac{\partial \hat{f}_2}{\partial x_2} & \frac{\partial \hat{f}_2}{\partial x_3} & ... & \frac{\partial \hat{f}_2}{\partial x_n} \\
  \frac{\partial \hat{f}_3}{\partial x_1} & \frac{\partial \hat{f}_3}{\partial x_2} & \frac{\partial \hat{f}_3}{\partial x_3} & ... & \frac{\partial \hat{f}_3}{\partial x_n} \\
  \vdots & \vdots & \vdots & \ddots & \vdots \\
  \frac{\partial \hat{f}_n}{\partial x_1} & \frac{\partial \hat{f}_n}{\partial x_2} & \frac{\partial \hat{f}_n}{\partial x_3} & ... & \frac{\partial \hat{f}_n}{\partial x_n} \\
\end{array}%
\right)-\left(%
\begin{array}{ccccc}
  p_1 & 0 & 0 & \cdots & 0 \\
  0 & p_2 & 0 & \cdots & 0 \\
  0 & 0 & p_3 & \cdots & 0 \\
  \vdots & \vdots & \vdots & \ddots & 0 \\
  0 & 0 & 0 & \cdots & p_n \\
\end{array}%
\right)\, ,
\end{equation}
at each non-zero balance $\vec{a}$ we found. Then the eigenvalues of the Kovalevskaya matrix should have the form $(-1,r_2,r_3,...,r_{n})$. If the eigenvalues $r_2,r_3,...,r_{n}$ are positive, then the dynamical system (\ref{dynamicalsystemdombalanceintro}) has general solutions which definitely lead to a finite-time singularity. This ensures the fact that all initial conditions lead to this general solution which in turn becomes singular at finite-time. If however one of $r_2,r_3,...,r_{n}$ is negative, then only a small set of initial conditions lead to a finite-time singularity, so further analysis is required, due to the fact that the solution is not general. In the case that a singularity is found, the singularity occurs at the orthant in the $R^n$ phase space which corresponds to the orthant of $\vec{a}$. This means that if for example $a_2$ is negative, then the singularity occurs at the orthant for which $x_2<0$. The proof of the theorem and details on the above results can be found in Ref. \cite{goriely}, see also \cite{barrowcotsakis}.

Let us here recapitulate in a few steps the dominant balance analysis which we shall employ in the next subsections, and after that we provide a simple example. We start with an autonomous $n-$dimensional polynomial dynamical system of the form
\begin{equation}\label{dynamicalsystemdombalanceintro}
\dot{x}=f(x)\, ,
\end{equation}
where $x$ is a real $n$-dimensional vector $x=(x_1,x_2,x_3,...,x_n)$ and also $f(x)=\left(f_1(x),f_2(x),...,f_n(x)\right )$, with each entry being a polynomial of the variables $x_1,x_2,x_3,...,x_n$. A finite-time singularity of the dynamical system (\ref{dynamicalsystemdombalanceintro}) is defined to be a singularity that strictly depends on the initial conditions of the dynamical system. The finite-time singularity at $t=c$ depends on the initial conditions chosen. Hence the method we shall describe in brief, actually determines the existence of such finite-time singularities in autonomous polynomial dynamical systems of the form (\ref{dynamicalsystemdombalanceintro}). The procedure has the following steps:

\begin{itemize}

\item We find truncations (decompositions) of the initial function $f(x)$ appearing in Eq. (\ref{dynamicalsystemdombalanceintro}), in certain dominant and also to subdominant parts, that is $f(x)=\hat{f}(x)+\breve{f}(x)$, where $\hat{f}(x)$ denotes the dominant part and $\breve{f}(x)$ denotes the subdominant part. The dominant part controls the behavior of the dynamical system near a finite time singularity, hence the dynamical system near a finite-time singularity becomes approximately,
\begin{equation}\label{dominantdynamicalsystem}
\dot{x}=\hat{f}(x)\, .
\end{equation}
In principle many truncations can be found.

\item Now for each variable $x_i$, $i=1,2,...,n$ of the vector $x$, we assume near the singularity that,
\begin{equation}\label{decompositionofxi}
x_1(\tau)=a_1\tau^{p_1},\,\,\,x_2(t)=a_2\tau^{p_2},\,\,\,....,x_n(t)=a_n\tau^{p_n}\, ,
\end{equation}
where $\tau=t-t_c$, and $t=t_c$ is the singularity time instance. Then for each truncation we substitute the $x_i$'s from in Eq. (\ref{decompositionofxi}) in Eq. (\ref{dominantdynamicalsystem}). By equating the polynomials obtained, we determine the parameters $p_i$, $i=1,2,...,n$. Only fractional or integer values are allowed for the parameters $p_i$. Then we form the vector $\vec{p}=(p_1,p_2,...,p_n)$.

\item Using the $p_i$'s from the previous step, we equate the polynomials in Eqs. (\ref{dominantdynamicalsystem}), (\ref{decompositionofxi}), and the coefficients $a_i$ are found. From these we form the vector $\vec{a}=(a_1,a_2,a_3,....,a_n)$.

\item The vectors $\vec{p}=(p_1,p_2,...,p_n)$ and $\vec{a}=(a_1,a_2,a_3,....,a_n)$ form a dominant balance $(\vec{a},\vec{p})$.

\item Having found the dominant balance $(\vec{a},\vec{p})$, the rest is simple algebra. If $\vec{a}=(a_1,a_2,a_3,....,a_n)$ takes complex values, then no finite-time singularities occur in the total dynamical system (\ref{dynamicalsystemdombalanceintro}). If it is real, then the dynamical system has singularities.

\item The next step is to determine whether these singularities correspond to global or local solutions. To be more precise if a general set of initial conditions leads to these singularities (global solutions), or if a limited set of initial conditions leads to these singularities (local solutions). This can be seen by the Kovalevskaya matrix $R$ in Eq. (\ref{kovaleskaya}). We calculate it for each non-zero $\vec{a}$ and we compute the eigenvalues which must have the form $(-1,r_2,r_3,...,r_{n})$.

\item If all the eigenvalues $r_2,r_3,...,r_{n}$ are positive, then the solution we found is general, and in all other cases, only local solutions exist. Specifically, if $\vec{a}$ contains some complex entries and if $r_i>0$, then no general singular solutions exist. If $\vec{a}$ has only real entries and $r_i>0$, then general singular solutions exist.

\end{itemize}

Before applying the method in the dynamical system (\ref{dynamicalsystemmultifluid}), it is worth providing a simple example in order to illustrate the simplicity of the method. We shall use an example from Ref. \cite{goriely}, so consider the following dynamical system,
\begin{equation}\label{example1}
\dot{x}_1=x_1(\alpha+bx_2),\,\,\,\dot{x}_2=cx_1^2+dx_2\, ,
\end{equation}
where $b,c>0$. The right hand side of the dynamical system corresponds to the two-dimensional vector $f(x_i)$, which is,
 \begin{equation}\label{totalf}
 f(x_i)=\left(%
\begin{array}{c}
  x_1(\alpha+b\,x_2) \\
 c\,x_1^2+dx_2 \\
\end{array}%
\right)\, ,
\end{equation}
so by applying the method we described above, the only dominant truncation $\hat{f}(x_i)$ that leads to a mathematically acceptable dominant balances $(\vec{a},\vec{p})$, is the following,
\begin{equation}\label{truncation}
 \hat{f}(x_i)=\left(%
\begin{array}{c}
  b\,x_1\,x_2 \\
 c\,x_1^2 \\
\end{array}%
\right)\, ,
\end{equation}
and only two balances are found, which we denote $(\vec{a}_1,\vec{p}_1)$ and $(\vec{a}_2,\vec{p}_1)$, where $\vec{a}_i$, $i=1,2$ and $\vec{p}_1$ are,
\begin{equation}\label{balancesexample1}
\vec{a}_1=(\frac{1}{\sqrt{b\,c}},\frac{1}{c}),\,\,\,\vec{a}_2=(-\frac{1}{\sqrt{b\,c}},\frac{1}{c}),\,\,\,p_1=(-1,-1)\, .
\end{equation}
By calculating the Kovalevskaya matrix $R$, we can easily see that the eigenvalues are $r_1=-1$ and $r_2=2$, so according to the theorem we described above, the autonomous dynamical system (\ref{balancesexample1}) has a general solution which ends up unavoidably to a finite-time singularity.

Hence the method is conceptually simple but it is mathematically rigid and may provide insights for cosmological systems that it is difficult, if not impossible, to solve these analytically. Of course, in order for the singularity analysis to make sense, the variables of the dynamical system must be chosen appropriately. As we discussed earlier, and we shall demonstrate in the next subsection, the dynamical system (\ref{dynamicalsystemmultifluid}), with the interaction term being chosen as in Eq. (\ref{additionalterms}), is an example of such a dynamical system, for which some physical outcomes may be obtained from the singularity analysis. In the next subsection we thoroughly investigate the singularity structure of the dynamical system (\ref{dynamicalsystemmultifluid}).

\subsection{Dominant Balance Analysis and Physical Interpretation of Singularities of the Dynamical System}

%%%%%%%%%%%%%%%%%%%%%%%%%%%%%%%%%rewrite in view of the Friedmann constraint. A singularity in x2 would imply the negative singularity in the other variable, the singularities must cancell due to the friedmann constraint!! Only x3 then should be positive.NEW THEORY AT HAND

% Dominant balances study, express x1 x2 and x3 as alpha and rho, define the balances, non trivial then use the following text

The structure of the variables (\ref{variablesofdynamicalsystem}) can reveal enough information for the occurrence of a finite-time singularity of the dynamical system. Indeed, if the variable $z$ is finite as a function of $N$, then $H^2$ is finite. In that case, if one of the $x_i$, $i=1,2,3$ diverges, then this would directly imply that one of the energy densities $\rho_d$, $\rho_m$ or $\rho_b$ would diverge, so this would imply a Big Rip or a Type II singularity loosely speaking, however things are not as simple as these initially look like. Firstly the variables $x_i$, $i=1,2,3$ satisfy the Friedmann constraint (\ref{friedmannconstraint}), so the singularities that occur in these variables must occur in such a way so that these eventually cancel, so that the Friedmann constraint is satisfied. So if the variable $z$ diverges, this would imply that if the $x_i$, $i=1,2,3$ diverge, the singularities should formally cancel. From a physical point of view, finding a singular baryon density is rather not physically appealing, so the only acceptable situation from a physical point of view is that the singularities occur in the dark sector variables $x_1$ and $x_2$, and these actually cancel. We saw in a previous section how this situation may be realized for the $f(R)$ gravity case, when Type IV singularities occur. Moreover, if  $x_1$ diverges, the EoS would diverge, but the important issue is that the leading order singularities in $x_1$ and $x_2$ should cancel, so for the balances, this would imply that $a_1=-a_2$ in the above equations, and also that the dominant exponent of the singularities for $x_1$ and $x_2$ should satisfy $p_1=p_2$.

In the case that $z$ is finite, if a singularity is found in the variables $x_i$, $i=1,2,3$, then the only physical acceptable case would also be the case in which $a_1=a_2$ in the above equations, and also $p_1=p_2$. Thus regardless if $z$ is finite or not, the singularities in the variables $x_i$, $i=1,2,3$ should eventually cancel. About the singularity type, this is not easy in general to answer, apart from the case that $z$ is singular. Loosely speaking, if the dynamical variables blow-up in the phase space as $N$ evolves, this could indicate possible regions of Big Rip or Type II singularities, however this is not necessarily true or obvious in general. For sure though, a singularity in the variable $z$ would strongly indicate a physical finite-time singularity, possibly a Type III case. Also in all cases, the variable $x_3$ must take positive values.

 In conclusion, the only case that a singularity is acceptable that can be verified for sure, is when the variable $z$ diverges at finite-time, and in turn this would indicate the possible presence of a finite-time singularity. Also a singularity in both $x_1$ and $x_2$ is also acceptable, but further caution and analysis is needed, since these must cancel. Suppose then that a singularity in $z$ is found, what would be the type of singularity? This crucially depends on the fact that a singularity in the Hubble rate expressed as a function of the cosmic time, may possibly imply a singularity in the corresponding $H(N)$. This can be seen by using the Hubble rate (\ref{hubblerate}), which in terms of the $e$-foldings number $N$ reads,
\begin{equation}\label{hubbleefoldingsfirstapprox}
H(N)\sim (N-N_c)^{\frac{\alpha}{\alpha-1}}\, .
\end{equation}
For the Big Rip case ($\alpha>1$), as we showed in the previous subsection, as $t\to t_s$, $N\to \infty$, hence indeed the Hubble rate diverges as the singularity $t\to t_s$ is approached. Also, the same applies for the Type III case, since recall that $0<\alpha<1$. For the Type II and Type IV cases, the corresponding Hubble rate $H(N)$ goes to zero. To our opinion then, only a singularity in the variable $z$ can surely indicate some possible Big Rip or Type III singularity physical singularity. So a mathematical singularity of the phase space in the variable $z$ is the only criterion that determines a physical singularity of the cosmological system. Notice that for the Big Rip and Type II cases, although that $H(N)$ diverge, the term $1/H(N)^2$ does not diverge but goes to zero as $N\to \infty$ for the Big Rip case, or as $N\to N_c$ for the Type III case. Hence the variables $x_i$ can be finite at the Big Rip or Type III singularities. Of course the Hubble rate (\ref{hubblerate}) is not a solution of the dynamical system, but we just used this example in order to find the variable responsible for a physical finite-time singularity of the cosmological system, which is $z$.

Therefore, a mathematical singularity of the dynamical system (\ref{dynamicalsystemmultifluid}) in one of the variables $x_1$, $x_2$ should be carefully interpreted, since one cannot know what sort of singularity this could be. However, due to the Friedmann constraint, the singularities in the variables $x_1$, $x_2$ must cancel, a condition which restricts the resulting form of the corresponding dominant balances. In Table \ref{table1}, we gathered the main conclusions of the above discussion, on how to interpret the singularity occurrence for the dynamical system (\ref{dynamicalsystemmultifluid}).
\begin{table*}[h]
\small \caption{\label{table1}Singularity Interpretation and Constraints for the Multifluid  Dynamical System (\ref{dynamicalsystemmultifluid})}
\begin{tabular}{@{}crrrrrrrrrrr@{}}
\tableline \tableline \tableline
 Singularity in $z$: Physical singularity, possibly a Big Rip or Type III.
\\\tableline
Singularity in $x_1$ and $x_2$: Acceptable singularities, but these must cancel,\\
 so the dominant balances must satisfy $a_1=-a_2$ and $p_1=p_2$.
\\\tableline
Singularity in $x_3$: Not physically acceptable.
\\\tableline
Constraints on $x_3$: It must be always positive and non-singular
\\\tableline
Interpretation of $x_1$ and $x_2$ singularities: If $z$ is singular, then possibly this indicates that $\rho_d$ and $\rho_m$ must diverge \\
in a way so that $\rho_m/H^2$ and $\rho_d/H^2$ is singular,
and these must satisfy $\rho_d=-\rho_m$. \\
If $z$ is regular, and also $x_1$ and $x_2$ are singular,
then this does not necessarily implies that $\rho_d$ and $\rho_m$ are singular,
\\
but certainly these must satisfy $\rho_d=-\rho_m$.
\\\tableline
In all cases, the Friedmann constraint $x_1+x_2+x_3=1$ must be satisfied, even at leading order.
\\\tableline
\tableline \tableline
\end{tabular}
\end{table*}
Before proceeding in finding all the dominant balances and to the corresponding analysis of the singularities, let us make some clarifications. We assume that the variables $x_1(N)$, $x_2(N)$, $x_3(N)$ and $z(N)$, near the possible singularities have the following leading order behavior,
\begin{equation}\label{decompositionofxiactualexample}
x_1(N)=a_1(N-N_c)^{p_1},\,\,\,x_2(N)=a_2(N-N_c)^{p_2},\,\,\, x_3(N)=a_3(N-N_c)^{p_3},\,\,\,z(N)=a_4(N-N_c)^{p_4}\, ,
\end{equation}
so we seek for balances of the form $(\vec{a},\vec{p})$, with $\vec{a}$ and $\vec{p}$ being of the form,
\begin{equation}\label{balancesactualcase}
\vec{a}=(a_1,a_2,a_3,a_4),\,\,\,\vec{p}=(p_1,p_2,p_3,p_4)\, .
\end{equation}
The dynamical system (\ref{dynamicalsystemmultifluid})  can be written as $\frac{\mathrm{d}\vec{x}}{\mathrm{d}N}=f(\vec{x})$, where $\vec{x}=(x_1,x_2,x_3,z)$, and the function $f(x_1,x_2,x_3,z)$ is,
\begin{equation}\label{functionfmultifluid}
f(x_1,x_2,x_3,z)=\left(%
\begin{array}{c}
 -c_1x_2-c_2x_1+9Ax_1^2z^2+3x_1x_2+3x_1x_3-9Azx_1^3 \\
 c_1x_2+c_2x_1-3x_2+3x_2^2+3x_2x_3-9Ax_1^2x_2z \\
  -3x_3+3x_3^2+3x_3x_2-9Ax_1^2x_3z \\
  -3x_2z-3x_3z+9Ax_1^2z^2 \\
\end{array}%
\right)
\end{equation}
So in the rest of this section, we shall seek for mathematically consistent truncations of the vector $f(x_1,x_2,x_3,z)$ appearing in Eq. (\ref{functionfmultifluid}), which will reveal the dominant behavior of the dynamical system near the singularities, if these exist, or will reveal that no singularities exist at all.

\subsection{First Mathematically Consistent Truncation}

The first mathematically consistent truncation of (\ref{functionfmultifluid}) is the following,
\begin{equation}\label{truncation1}
\hat{f}(x_1,x_2,x_3,z)=\left(
\begin{array}{c}
 9 A x_1(N)^2 z(N)^2 \\
 3 x_2(N)^2 \\
 3 x_3(N)^2 \\
 9 A x_1(N)^2 z(N)^2 \\
\end{array}
\right)\, ,
\end{equation}
so by following the procedure we presented in the previous subsection, the following solution for the vector $\vec{p}$ is obtained,
\begin{equation}\label{vecp1}
\vec{p}=( -\frac{1}{3}, -1, -1, -\frac{1}{3} )
\end{equation}
Accordingly, for $\vec{p}$ being as above, the following vector-solutions $\vec{a}_1$, $\vec{a}_2$ and $\vec{a}_3$ are found,
\begin{align}\label{balancesdetails1}
& \vec{a}_1=\Big{(}-\frac {1} {3 A^{1/3}}, -\frac {1} {3}, -\frac {1} {3}, -\frac {1} {3 A^{1/3}}\Big{)} \\ \notag &
\vec{a}_2=\Big{(}\frac {(-1)^{1/
     3}} {3 A^{1/3}}, -\frac {1} {3}, -\frac {1} {3}, \frac {(-1)^{1/
     3}} {3 A^{1/3}}\Big{)} \\ \notag &
\vec{a}_3=\Big{(}-\frac {(-1)^{1/
     3}} {3 A^{1/3}}, -\frac {1} {3}, -\frac {1} {3}, -\frac {(-1)^{1/
     3}} {3 A^{1/3}}\Big{)} \, ,
\end{align}
and the Kovalevskaya matrix for the truncation (\ref{truncation1}) is,
\begin{equation}\label{kobvalev1}
R_1=\left(
\begin{array}{cccc}
 18 A x_1 z^2+\frac{1}{3} & 0 & 0 & 18 A x_1^2 z \\
 0 & 6 x_2+1 & 0 & 0 \\
 0 & 0 & 6 x_3+\frac{2}{3} & 0 \\
 18 A x_1 z^2 & 0 & 0 & 18 A z x_1^2+\frac{1}{3} \\
\end{array}
\right)\, .
\end{equation}
Let us find the Kovalevskaya matrix for each $\vec{a}_i$, $i=1,2,3$ we found in Eq. (\ref{balancesdetails1}), so for $\vec{a}_1$, the Kovalevskaya matrix becomes,
\begin{equation}\label{kovalevskayacase1}
R_1(\vec{a}_1)=\left(
\begin{array}{cccc}
 -\frac{1}{3} & 0 & 0 & -\frac{2}{3} \\
 0 & -1 & 0 & 0 \\
 0 & 0 & -\frac{4}{3} & 0 \\
 -\frac{2}{3} & 0 & 0 & -\frac{1}{3} \\
\end{array}
\right)\, ,
\end{equation}
and the corresponding eigenvalues for $\vec{a}_1$ are,
\begin{equation}\label{eigenvalues1}
(r_1,r_2,r_3,r_4)=(-\frac{4}{3},-1,-1,\frac{1}{3})\, .
\end{equation}
For $\vec{a}_2$ and $\vec{a}_3$, the Kovalevskaya matrix is the same as for the vector $\vec{a}_1$, and the corresponding eigenvalues are the same. So the truncation (\ref{truncation1}) does not lead to mathematically appealing results, since in all the cases the first eigenvalue is $r_1\neq -1$. Thus no conclusion can be made for the singularity structure in this case.

\subsection{Second Mathematically Consistent Truncation}

The second mathematically consistent truncation of (\ref{functionfmultifluid}) that can be formed, is the following,
\begin{equation}\label{truncation12}
\hat{f}(x_1,x_2,x_3,z)=\left(
\begin{array}{c}
 9 A x_1(N)^2 z(N)^2 \\
 3 x_2(N) x_3(N) \\
 3 x_2(N) x_3(N) \\
 9 A x_1(N)^2 z(N)^2 \\
\end{array}
\right)\, ,
\end{equation}
which leads to the following $\vec{p}$,
\begin{equation}\label{vecp12}
\vec{p}=( -\frac{1}{3}, -1, -1, -\frac{1}{3} )
\end{equation}
and the corresponding vectors  $\vec{a}_1$, $\vec{a}_2$ and $\vec{a}_3$, are equal to,
\begin{align}\label{balancesdetails12}
& \vec{a}_1=\Big{(}-\frac {1} {3 A^{1/3}}, -\frac {1} {3}, -\frac {1} {3}, -\frac {1} {3 A^{1/3}}\Big{)} \\ \notag &
\vec{a}_2=\Big{(}\frac {(-1)^{1/
     3}} {3 A^{1/3}}, -\frac {1} {3}, -\frac {1} {3}, \frac {(-1)^{1/
     3}} {3 A^{1/3}}\Big{)} \\ \notag &
\vec{a}_3=\Big{(}-\frac {(-1)^{2/
     3}} {3 A^{1/3}}, -\frac {1} {3}, -\frac {1} {3}, -\frac {(-1)^{2/
     3}} {3 A^{1/3}}\Big{)} \, .
\end{align}
Accordingly, for the truncation (\ref{truncation12}), the Kovalevskaya matrix is,
\begin{equation}\label{kobvalev12}
R_1=\left(
\begin{array}{cccc}
 18 A x_1 z^2+\frac{1}{3} & 0 & 0 & 18 A x_1^2 z \\
 0 & 3 x_3+1 & 3 x_2 & 0 \\
 0 & 3 x_3 & 3 x_2+1 & 0 \\
 18 A x_1 z^2 & 0 & 0 & 18 A z x_1^2+\frac{1}{3} \\
\end{array}
\right)\, ,
\end{equation}
For $\vec{a}_1$, the Kovalevskaya matrix becomes,
\begin{equation}\label{kovalevskayacase12}
R_1(\vec{a}_1)=\left(
\begin{array}{cccc}
 -\frac{1}{3} & 0 & 0 & -\frac{2}{3} \\
 0 & 0 & -1 & 0 \\
 0 & -1 & 0 & 0 \\
 -\frac{2}{3} & 0 & 0 & -\frac{1}{3} \\
\end{array}
\right)\, ,
\end{equation}
and the corresponding eigenvalues are,
\begin{equation}\label{eigenvalues12}
(r_1,r_2,r_3,r_4)=(-1,-1,1,\frac{1}{3})
\end{equation}
For $\vec{a}_2$, the Kovalevskaya matrix is,
\begin{equation}\label{kovalevskayacase12tispopis}
R_1(\vec{a}_1)=\left(
\begin{array}{cccc}
 1 & 0 & 0 & \frac{2}{3} \\
 0 & 0 & -1 & 0 \\
 0 & -1 & 0 & 0 \\
 \frac{2}{3} & 0 & 0 & 1 \\
\end{array}
\right)\, ,
\end{equation}
and the corresponding eigenvalues are,
\begin{equation}\label{eigenvalues12extracase2}
(r_1,r_2,r_3,r_4)=(\frac{5}{3},-1,1,\frac{1}{3})\, ,
\end{equation}
Finally, for $\vec{a}_3$, the Kovalevskaya matrix is identical to the one appearing in Eq. (\ref{kovalevskayacase12}) and the eigenvalues are given in Eq. (\ref{eigenvalues12}). So the only cases of interest correspond to $\vec{a}_1$ and $\vec{a}_3$, for which the first eigenvalue is $r_1=-1$, so the case $\vec{a}_2$ is ruled out.

\subsubsection{Physical Analysis of the Results for the Second Truncation}

For the cases of interest, namely those which correspond to $\vec{a}_1$ and $\vec{a}_3$, we now analyze the occurrence of singularities in the phase space of the dynamical system (\ref{dynamicalsystemmultifluid}). For both $\vec{a}_1$ and $\vec{a}_3$, the whole singularity interpretation crucially depends on the sign of the parameter $A$. If $A$ is positive, then $\vec{a}_1$ has real components and $\vec{a}_3$ has always complex components. On the other hand, in the case $A<0$, the whole analysis depends crucially on the interpretation of the term containing $(-1)^{1/3}$ in $\vec{a}_1$, since $\vec{a}_3$ has real components. Indeed, the term $(-1)^{1/3}$, corresponds to the solution of the equation $x^3=-1$, which has three roots, one real, $x=-1$ and two complex conjugate $x_{1,2}=0.5\pm 0.866025$. So if we keep the real root of $(-1)^{1/3}$, $\vec{a}_1$ has real components, and therefore due to the fact that $\vec{a}_3$ is also real, this indicates that the dynamical system (\ref{dynamicalsystemmultifluid}) has finite-time singularities, and we need to investigate if a general solution that leads the dynamical system to the singularities exists, or the result holds true only for a small set of initial conditions. This can be seen from the eigenvalues (\ref{eigenvalues12}), so due to the fact that $r_2<0$, this means that the solution is not general and only a limited set of initial conditions lead to singularities. In addition, the singularity structure for the variables $x_1$, $x_2$ and $x_3$ does not have the desired form, as it can be seen from the structure of the vector $\vec{p}$ appearing in Eq. (\ref{vecp1}). This behavior occurs due to the fact that the leading order exponents are not of the same order for $x_1$, $x_2$ and $x_3$, so the Friedmann constraint (\ref{friedmannconstraint}) is not satisfied. Therefore, the small set of initial conditions that lead to singularities do not yield physically acceptable results. Let us proceed to the case that $A<0$ while we keep the complex roots of $(-1)^{1/3}$. In this case $\vec{a}_3$ is still real, however $\vec{a}_1$ is complex, therefore this means that no singularities occur in the dynamical system, however in this case the values of the eigenvalues (\ref{eigenvalues12}) reveals that the solution is not general, and only a small set leads the solutions of the dynamical system to a non-singular region. Nevertheless, these non-singular solutions are physically acceptable, in contrast to the singular solutions case.

Hence for the truncation (\ref{truncation12}), the dynamical system does not have general solutions which drive the dynamics to a finite-time singularity.

\begin{figure}[h]
\centering
\includegraphics[width=18pc]{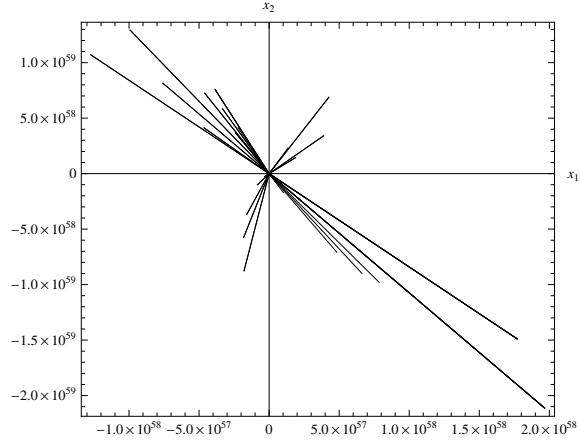}
\caption{{\it{The trajectories in the $x_1-x_2$ plane for the multifluid cosmological dynamical system and for various initial conditions.}}} \label{finalplots1newa}
\end{figure}

\subsection{Third Mathematically Consistent Truncation}

A third mathematically consistent truncation of (\ref{functionfmultifluid}) is the following,
\begin{equation}\label{truncation13}
\hat{f}(x_1,x_2,x_3,z)=\left(
\begin{array}{c}
 9 A x_1(N)^2 z(N)^2 \\
 3 c_2 x_1(N) \\
 3 x_3(N)^2 \\
 9 A x_1(N)^2 z(N)^2 \\
\end{array}
\right)\, ,
\end{equation}
which leads to the following $\vec{p}$,
\begin{equation}\label{vecp13}
\vec{p}=( -\frac{1}{3}, \frac{2}{3}, -1, -\frac{1}{3} )
\end{equation}
and the corresponding vectors $\vec{a}_i$, $i=1,2,3$, are in this case,
\begin{align}\label{balancesdetails13}
& \vec{a}_1=\Big{(}-\frac {1} {3 A^{1/3}}, -\frac{3 c_2}{2 A^{1/3}}, -\frac {1} {3}, -\frac {1} {3 A^{1/3}}\Big{)} \\ \notag &
\vec{a}_2=\Big{(}\frac {(-1)^{2/
     3}} {3 A^{1/3}}, -\frac{3 c_2(-1)^{2/
     3}}{2 A^{1/3}}, -\frac {1} {3}, \frac {(-1)^{2/
     3}} {3 A^{1/3}}\Big{)} \\ \notag &
\vec{a}_3=\Big{(}-\frac {(-1)^{2/
     3}} {3 A^{1/3}}, -\frac{3 c_2(-1)^{2/
     3}}{2 A^{1/3}}, -\frac {1} {3}, -\frac {(-1)^{2/
     3}} {3 A^{1/3}}\Big{)} \, ,
\end{align}
\begin{figure}[h]
\centering
\includegraphics[width=18pc]{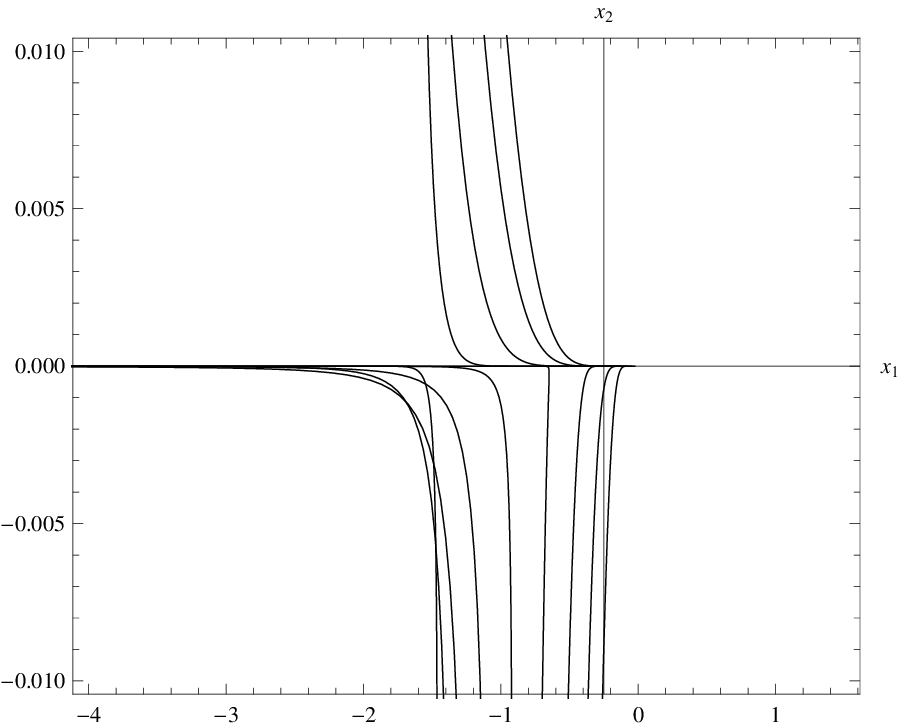}
\caption{{\it{The trajectories in the $x_1-x_2$ plane for the multifluid cosmological dynamical system and for various initial conditions.}}} \label{finalplots1newb}
\end{figure}
while the Kovalevskaya matrix is,
\begin{equation}\label{kobvalev13}
R_1=\left(
\begin{array}{cccc}
 18 A x_1 z^2+\frac{1}{3} & 0 & 0 & 18 A x_1^2 z \\
 3 c_2 & -\frac{2}{3} & 0 & 0 \\
 0 & 0 & 6 x_3+1 & 0 \\
 18 A x_1 z^2 & 0 & 0 & 18 A z x_1^2+\frac{1}{3} \\
\end{array}
\right)\, ,
\end{equation}
For $\vec{a}_1$, the Kovalevskaya matrix becomes,
\begin{equation}\label{kovalevskayacase13}
R_1(\vec{a}_1)=\left(
\begin{array}{cccc}
 -\frac{1}{3} & 0 & 0 & -\frac{2}{3} \\
 3 c_2 & -\frac{2}{3} & 0 & 0 \\
 0 & 0 & -1 & 0 \\
 -\frac{2}{3} & 0 & 0 & -\frac{1}{3} \\
\end{array}
\right)\, ,
\end{equation}
and the corresponding eigenvalues are,
\begin{equation}\label{eigenvalues13}
(r_1,r_2,r_3,r_4)=(-1,-1,-\frac{2}{3},\frac{1}{3})
\end{equation}
For $\vec{a}_2$ and $\vec{a}_3$, the Kovalevskaya matrix is the same as for the balance $\vec{a}_1$, and the eigenvalues are the same. Thus this is a degenerate situation, so let us analyze the $\vec{a}_1$ case. The physical analysis of singularities occurrence is similar with the truncation (\ref{truncation12}), so the resulting picture depends on the sign of $A$ and the roots of $(-1)^{1/3}$. Hence, the dynamical system for the truncation (\ref{truncation13}) does not have general solutions which lead to finite-time singularities, or to put more correctly, there is a limited set of initial conditions which leads the dynamical system to finite-time singularities, which are unphysical though, due to the form of $\vec{p}$ in this case.

In conclusion, the resulting picture after the dominant balance analysis for all possible balances is the following: The dynamical system does not have general solutions which lead to physical singularities. Of course, this does not exclude the possibility of having some initial conditions that may lead to singularities, but in this case the singularities are unphysical, as we demonstrated above. However this conclusion is entirely based on the dominant balance analysis, which works only for vectors $\vec{a}_i\neq 0$. Therefore it is rather compelling to analyze the phase space of the dynamical system (\ref{dynamicalsystemmultifluid}) in order to see whether the behavior of the trajectories explicitly. This is the subject of the next subsection.

\begin{figure}[h]
\centering
\includegraphics[width=20pc]{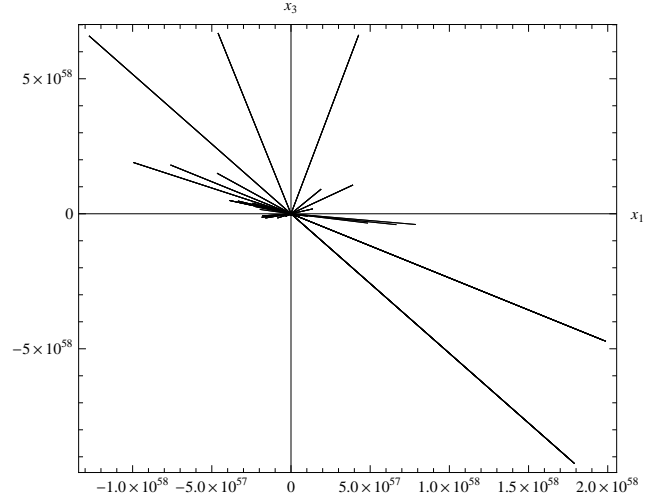}
\caption{{\it{The trajectories in the $x_1-x_3$ plane for the multifluid cosmological dynamical system and for various initial conditions.}}} \label{finalplots2newa}
\end{figure}

\subsection{Phase Space Analysis for the Multifluid Cosmological Model}

Our previous analysis showed that there exist some initial conditions which generate solutions which become eventually singular. Also we showed that there exist also initial conditions which may lead to non-singular solutions, with non-singular meaning that no finite-time singularity occurs. However no general solution was found, that is singular or non-singular. In order to have a more complete picture of the phase space of the dynamical system (\ref{dynamicalsystemmultifluid}), in this subsection we shall study in some detail the structure of the phase space and the corresponding trajectories of the dynamical system (\ref{dynamicalsystemmultifluid}).
\begin{figure}[h]
\centering
\includegraphics[width=20pc]{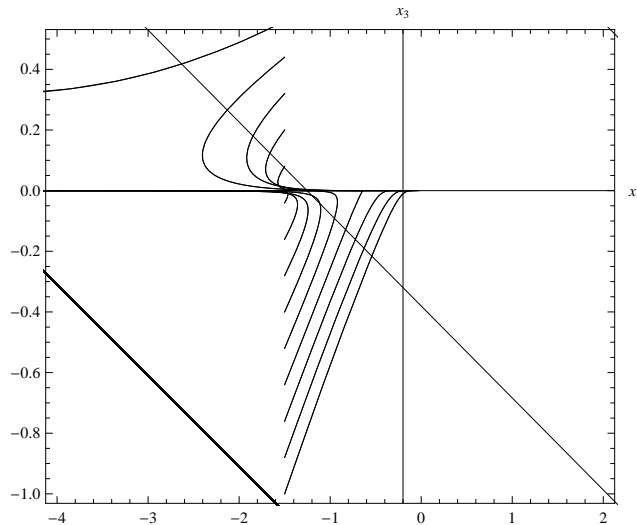}
\caption{{\it{The trajectories in the $x_1-x_3$ plane for the multifluid cosmological dynamical system and for various initial conditions.}}} \label{finalplots2newb}
\end{figure}
Particularly, we shall be interested in finding the fixed points of the system and we examine their stability towards linear perturbations, by using the Hartman-Grobman theorem, if it applies. Also we shall perform some numerical analysis of the dynamical system at hand and we shall present some characteristic trajectories corresponding to various initial conditions. It is worth recalling some fundamental features of the theoretical framework of dynamical systems analysis, and for more details we refer the reader to Refs. \cite{jost,wiggins,voyatzis}.
\begin{figure}[h]
\centering
\includegraphics[width=18pc]{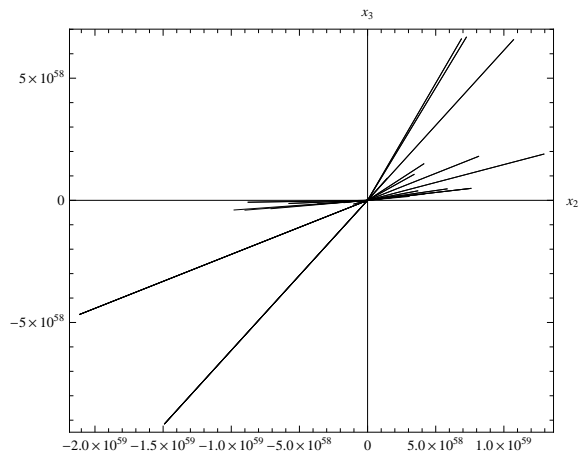}
\caption{{\it{The trajectories in the $x_2-x_3$ plane for the multifluid cosmological dynamical system and for various initial conditions.}}} \label{finalplots3newa}
\end{figure}
The standard approach in studying and analyzing non-linear dynamical systems is based on the linearization procedure and particularly on the Hartman-Grobman linearization theorem \cite{wiggins}. The Hartman-Grobman linearization theorem actually determines the stability of a fixed point and reveals the topological structure of the phase space of the full non-linear dynamical system, in the case of course that the fixed points of the dynamical system are hyperbolic.
\begin{figure}[h]
\centering
\includegraphics[width=18pc]{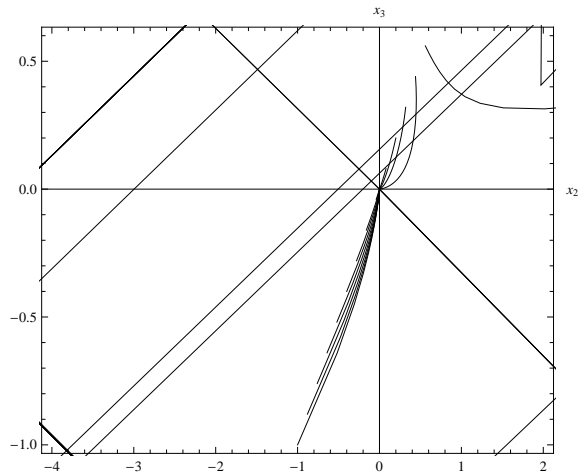}
\caption{{\it{The trajectories in the $x_2-x_3$ plane for the multifluid cosmological dynamical system and for various initial conditions.}}} \label{finalplots3newb}
\end{figure}
In the following we shall briefly recall these issues, since these will be useful in the analysis that follows. Let the vector field  $\Phi (t)$
$\epsilon$ $R^n$ satisfy the following differential flow,
\begin{equation}\label{ds1}
\frac{\mathrm{d}\Phi}{\mathrm{d}t}=g(\Phi (t))\, ,
\end{equation}
where $g(\Phi (t))$  is a locally Lipschitz continuous map of the form $g:R^n\rightarrow R^n$. We denote the fixed points of the dynamical system (\ref{ds1}) as $\phi_*$ and also with $\mathcal{J}(g)$ we denote the Jacobian matrix which corresponds to the linearized dynamical system near the fixed point, and it is equal to,
\begin{equation}\label{jaconiab}
\mathcal{J}=\sum_i\sum_j\Big{[}\frac{\mathrm{\partial f_i}}{\partial
x_j}\Big{]}\, .
\end{equation}
The Jacobian matrix is evaluated at the fixed points, and its eigenvalues will reveal the stability of the fixed point, only if the fixed point is hyperbolic. Recall that a fixed point is hyperbolic only if the spectrum $\sigma
(\mathcal{J})$ of the eigenvalues of the Jacobian matrix, consists of elements $e_i$ which satisfy $\mathrm{Re}(e_i)\neq 0$. Then, according to the Hartman-Grobman theorem, the linearized dynamical system,
\begin{equation}\label{loveisalie}
\frac{\mathrm{d}\Phi}{\mathrm{d}t}=\mathcal{J}(g)(\Phi)\Big{|}_{\Phi=\phi_*}
(\Phi-\phi_*)\, ,
\end{equation}
is topologically equivalent to the dynamical system (\ref{ds1}) near the hyperbolic fixed points $\phi_*$. Particularly, the theorem indicates that there exists a homeomorphism  $\mathcal{F}:U\rightarrow R^n$ in an open neighborhood $U$ of a hyperbolic fixed point $\phi_*$, which generates a flow $\frac{\mathrm{d}h(u)}{\mathrm{d}t}$, which satisfies,
\begin{equation}\label{fklow}
\frac{\mathrm{d}h(u)}{\mathrm{d}t}=\mathcal{J}h(u)\, .
\end{equation}
The flows (\ref{fklow}) and (\ref{ds1}) are homeomorphic. With regard to the stability of each fixed point $\phi_*$, if all the eigenvalues of the Jacobian matrix satisfy $\mathrm{Re}\left(\sigma (\mathcal{J}(g))\right)<0$, then the fixed point is asymptotically stable. In all other cases, the fixed point is unstable.

Let us now focus on the dynamical system (\ref{dynamicalsystemmultifluid}) for the interaction term chosen as in Eq. (\ref{qtermform}), and we shall try to find the fixed points of the system and study their stability towards linear perturbations, by applying the Hartman-Grobman theorem. As it turns out, for general and non-zero values of the parameter $c_2$, the analytic form of the fixed points is particularly complicated and extended that it is impossible to quote here. So in the following we shall divide the study in two cases. In the first case we shall analyze numerically the behavior of the fixed points, and in the second case we shall set $c_2=0$. In the latter case, the interaction term containing the energy density of the dark energy sector is turned off, so only the dark matter energy density remains active as interaction term, and the fixed points acquire a simplified closed form. Therefore, the stability analysis proves to be quite easy. Interactions of the form (\ref{qtermform}), with $c_2=0$ are also quite frequently used in the literature, see for example \cite{Boehmer:2008av}.

Let us start with the case in which both the terms proportional to $c_1$ and $c_2$ are active in the interaction term (\ref{qtermform}). We performed a detailed numerical analysis of the dynamical system, for four different cases, which we classify in Table \ref{numericalanalysisofsystem}. As it proves, in all cases, namely cases I-IV in Table \ref{numericalanalysisofsystem}, the various different and with physical significance fixed points in each case are unstable. Thus the general $c_1$ and $c_2$ case indicates instability in the dynamical system, regardless the value of the parameter $A$. \begin{table*}[h]
\small \caption{\label{numericalanalysisofsystem}Stability Results of the Fixed Points for the Multifluid  Dynamical System (\ref{dynamicalsystemmultifluid}) for General $c_1$ and $c_2$.}
\begin{tabular}{@{}crrrrrrrrrrr@{}}
\tableline \tableline \tableline
 Case I: $c_1>0$, $c_2>0$, for all $A$:  Unstable Fixed Points.
\\\tableline
Case II: $c_1>0$, $c_2<0$, for all $A$:  Unstable Fixed Points.
\\\tableline
Case III: $c_1<0$, $c_2>0$, for all $A$:  Unstable Fixed Points.
\\\tableline
Case IV: $c_1<0$, $c_2<0$, for all $A$:  Unstable Fixed Points.
\\\tableline
\tableline \tableline
\end{tabular}
\end{table*}
The instability of the fixed points persists even in the case that $c_2=0$ as we now demonstrate analytically. By setting $c_2=0$ the fixed points of the dynamical system (\ref{dynamicalsystemmultifluid}) acquire a simple form and these are,
\begin{align}\label{fixedpointsc20}
& \phi_1=\{x_1\to 0,x_2\to 0,x_3\to 0\}, \\ \notag &
\phi_2=\{x_2\to 0,x_3\to 0,z\to 0\},\\ \notag &
\phi_3=\{x_1\to 0,x_2\to 0,x_3\to 1,z\to 0\},\\ \notag &
\phi_4=\{x_1\to c_1,x_2\to 1-c_1,x_3\to 0,z\to 0\},\\ \notag &
\phi_5=\left\{x_1\to \frac{3 A c_1-\sqrt{9 A^2 c_1^2-12 A (c_1-1) c_1}}{6 A c_1},x_2\to \frac{1}{2} \left(3 A c_1-\sqrt{3} \sqrt{A c_1 (3 A c_1-4 c_1+4)}-2 c_1+2\right),x_3\to 0,z\to c_1\right\}\\ \notag &
\phi_6=\left.x_1\to \frac{\sqrt{9 A^2 c_1^2-12 A (c_1-1) c_1}+3 A c_1}{6 A c_1},x_2\to \frac{1}{2} \left(\sqrt{9 A^2 c_1^2-12 A (c_1-1) c_1}+3 A c_1-2 c_1+2\right),x_3\to 0,z\to c_1\right\}\, .
\end{align}
The Jacobian matrix $\mathcal{J}$ in the case at hand is,
\begin{align}\label{jacobianmatrix}
& \mathcal{J}=\\ \notag & \left(
\begin{array}{cccc}
 3 x_2+3 x_3+9 A x_1 z (2 z-3 x_1) & 3 x_1-3 c_1 & 3 x_1 & -9 A x_1^2 (x_1-2 z) \\
 -9 A x_2 z & 3 (c_1+2 x_2+x_3-3 A x_1 z-1) & 3 x_2 & -9 A x_1 x_2 \\
 -18 A x_1 x_3 z & 3 x_3 & 3 \left(-3 A z x_1^2+x_2+2 x_3-1\right) & -9 A x_1^2 x_3 \\
 18 A x_1 z^2 & -3 z & -3 z & -3 \left(-6 A z x_1^2+x_2+x_3\right) \\
\end{array}
\right)\, .
\end{align}
\begin{figure}[h]
\centering
\includegraphics[width=18pc]{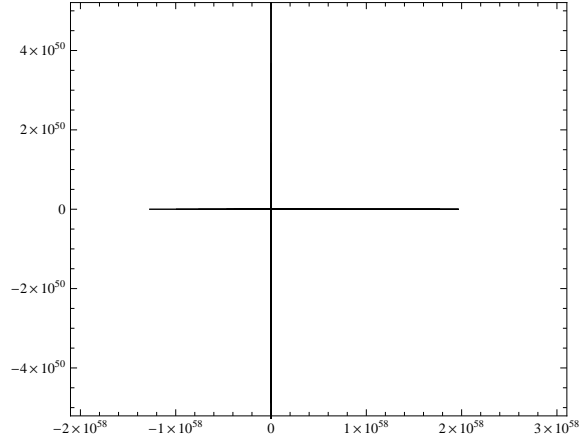}
\caption{{\it{The trajectories in the $x_1-z$ plane for the multifluid cosmological dynamical system and for various initial conditions.}}} \label{finalplots4newa}
\end{figure}
In the last two cases, the fixed points lead to de Sitter vacua, since $z=c_1$ at the fixed point. In all other cases, $z$ is zero at the fixed point, so the most interesting cases phenomenologically correspond to the last two fixed points, namely $\phi_5$ and $\phi_6$, in which case $x_3$ tends to zero. This means that asymptotically near the fixed point the baryonic fluid has no contribution to the dynamical evolution of the Universe near the two fixed points. Also, for consistency reasons, $c_1$ must be positive. The general form of the resulting eigenvalues of the Jacobian matrix, for the first four fixed points is the following,
\begin{align}\label{firsfoureigenvalues}
& \phi^*_1\to \{-3,0,0,3 c_1-3\},\\ \notag &
\phi^*_2\to \{-3,0,0,3 c_1-3\}, \\ \notag &
\phi^*_3 \to \{-3,3,3,3 c_1\}, \\ \notag &
\phi^*_4 \to \{3 (1-c_1)-3,-3 (1-c_1),3 (1-c_1),3 (1-c_1)\}\, .
\end{align}
The first two fixed points are not hyperbolic, so the Hartman-Grobman theorem does not apply in this case. However, the next two fixed points, namely $\phi^*_3$ and $\phi^*_4$ are hyperbolic ones, however due to the fact that some eigenvalues are positive, this means that these fixed points are not stable fixed points. So the first four fixed points are unstable fixed points, however not so physically appealing since these lead to a Hubble rate that goes to zero.
\begin{figure}[h]
\centering
\includegraphics[width=18pc]{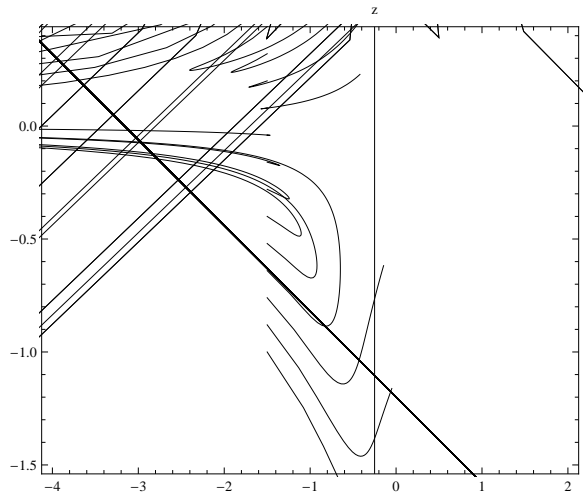}
\caption{{\it{The trajectories in the $x_1-z$ plane for the multifluid cosmological dynamical system and for various initial conditions.}}} \label{finalplots4newb}
\end{figure}
Let us now focus on the last two fixed points, namely $\phi^*_5$ and $\phi^*_6$. In this case, the determination of the eigenvalues ends up to a complicated algebraic equation which is rather inconvenient to be presented here. A thorough numerical analysis though, for $c_1>0$ and for various positive and negative values of $A$, shows that in this case too the fixed points are strictly unstable hyperbolic de Sitter fixed points. The source of instability will be analyzed in another work since this is beyond the purposes of this work, but let us briefly mention that the presence of the term proportional to $A$ in the EoS of dark energy and also due to the fact that the linear term $\rho$ in the EoS appears as $-\rho$ and not as $w_d\rho$, the instability of the dynamical system occurs.

Let us now examine the behavior of the trajectories for this dynamical system, in order to see at first hand if the behavior we found in the previous subsection is supported by the phase space trajectories. A detailed analysis of the phase space for a wide range of the parameters $A$ and $c_1$, shows that the analysis we performed in the previous subsection can also be seen in the phase space trajectories. Indeed, as we will now demonstrate by using some illustrative phase space plots, there are initial conditions for the dynamical system which lead to finite-time blow-ups in the dynamical system, and other trajectories that lead to non-singular behaviors.

In Figs. \ref{finalplots1newa} and \ref{finalplots1newb} we present the phase space trajectories in the $x_1-x_2$ plane, and accordingly in Figs. \ref{finalplots2newa} and \ref{finalplots2newb}, Figs. \ref{finalplots3newa} and \ref{finalplots3newb} and finally in Figs. \ref{finalplots4newa} and \ref{finalplots4newb} the trajectories in the planes $x_1-x_3$, $x_2-x_3$ and $x_1-z$ respectively. We omitted for brevity the plots for the plane $x_2-z$ since the behavior is exactly the same as in the $x_1-z$ plane. Actually, as it can be seen from Fig. \ref{finalplots4newb}, the behavior looks quite complicated near the origin. Furthermore, there exist highly singular solutions corresponding to several initial conditions, that blow up, see for example Fig. \ref{finalplots4newa} in which we present two of these. The plot is quite simple since the trajectories near the origin are practically insignificant in comparison to the plotted singular trajectories. Also, in all the plots it can be seen that there are regular and singular solutions, but these strongly depend on the initial conditions and these are not global attractor solutions. This is also supported from the instability of the fixed points. Of course this behavior occurs for the particular dynamical system at hand, and we aim to extend the physical EoS of the dark energy sector in order to see whether this behavior persists. In addition, the same analysis should be performed in the context of Loop Quantum Cosmology, after appropriately constructing the dynamical system. Such a construction is possible and we aim to report in a future work the results of the corresponding analysis.

In conclusion, the dynamical system (\ref{dynamicalsystemmultifluid}) corresponding to the interacting multifluid cosmology, is unstable and no global attractor solution exists that may lead to finite-time blow-up singularities, only some solutions which correspond to a small set of initial conditions.

\section{Conclusions}

In this work we thoroughly studied finite-time cosmological singularities from the dynamical system perspective of the cosmological system. Due to the fact that the finite-time singularities occurring in the variables of the dynamical system, do not necessarily imply the existence of a cosmological finite-time singularity, we divided the study in cosmological systems corresponding to a characteristic theory of modified gravity, namely $f(R)$ gravity in vacuum, and in cosmological systems of interacting fluids. In the $f(R)$ gravity case, the variables we used do not reveal the finite-time singularity structure of the cosmological system, even if these blow-up in the phase space. In effect these theories require a special treatment, so we investigated the behavior of vacuum $f(R)$ gravity near a finite-time singularity of the simplest form. In this case, the dynamical system is rendered non-autonomous, however due to the fact that we constructed the dynamical system in a particularly convenient form, we were able to integrate the equations analytically, for some limiting cases of the parameter $\alpha$, and we revealed the behavior of vacuum $f(R)$ gravity near a Big Rip and Type IV cosmological finite-time singularities. As we explicitly demonstrated, some of the variables of the dynamical system do not blow-up near a finite-time cosmological singularity, so this feature verified our claim about the difference between finite-time cosmological singularities and finite-time singularities of a dynamical system. For the $f(R)$ gravity case, we studied analytically only the Big Rip, and the Type IV singularities. The Big Rip case offered some exciting findings, due to the fact that in this case the initial non-autonomous dynamical system is rendered autonomous asymptotically near the Big Rip singularity. The most important finding in this case is that the fixed point of the asymptotically autonomous dynamical system is also the attractor of the non-autonomous dynamical system. And due to the fact that the fixed point of the asymptotically autonomous dynamical system describes an accelerated expansion with EoS $w_{eff}=-1$, we argued that a Big Rip singularity is always related to an accelerated expansion, at least in the context of $f(R)$ gravity and when the singularity is extremely strong. Also from a mathematical point of view, our results are interesting since we proved that in the Big Rip singularity, the asymptotically autonomous and the non-autonomous dynamical system share the same attractor. We aim to further develop and study this feature in a future work.

The second part of this work was devoted in the cosmological dynamical system corresponding to a multifluid Universe. Particularly, we assumed that the Universe consists of three viscous in general fluids, the baryon fluid, the dark energy and the dark matter fluid. Also we assumed that the dark sector fluids interact with each other, and also that these do not interact at all with the baryonic fluid. After choosing a particular form of the interaction term and also of the dynamical system variables, we formed a polynomial autonomous dynamical system and we investigated the finite-time singularity structure of the dynamical system. Due to the fact that the variables contain the Hubble rate and the energy densities of the fluids, the finite-time singularities of the dynamical system can reveal some information for the cosmological singularity structure of the cosmological system. We thoroughly discussed when a finite-time singularity of the dynamical system may imply a cosmological singularity for the cosmological system, and by using a rigid mathematical theorem which applies to autonomous polynomial dynamical systems, we investigated if finite-time singularities occur for the dynamical system. Our analysis showed than the dynamical system does not have a general solution leading to cosmological singularities, however there exists a small set in the phase space that may lead to finite-time singularities, however these are unphysical. Also we performed a numerical and analytical analysis of the resulting dynamical system.

In principle, our study may be generalized by using different equations of state for the dark energy fluid, or if the whole framework is examined from the Loop Quantum Cosmology perspective. More importantly, the baryonic fluid can be disregarded in all the aforementioned studies. We have strong hints that the presence of the extra baryonic fluid affects crucially the singularity structure of the cosmological system and certainly obscures the resulting picture of the phase space. Actually, the absence of such fluid may lead to general solutions which lead to finite-time singularities, or to general non-singular solutions.

An interesting application of the dominant balance singularity analysis which we used in this work, can possibly be applied in the system of cosmological equations corresponding to compact stars such as neutron stars. In fact, the singularity development in this kind of systems was firstly addressed in Ref. \cite{Bamba:2011sm}, so a direct application is to apply the dominant balance analysis in the resulting singularity evolution differential equation in order to see at first hand, whether a singularity can occur. Also, the astrophysical equations for a compact star in Einstein gravity or $f(R)$ gravity can be formed as a dynamical system of polynomial type, if the variables are appropriately chosen, so the dominant balance analysis can be directly applied in this case.

Finally, an interesting task would be to find the corresponding picture in the Einstein frame dynamical system, by conformally relating the two dynamical systems. This task is challenging from a mathematical point of view, since the Einstein and Jordan frame dynamical systems might be related. However, the physical significance of the conformally transformed dynamical system should be crucially examined for consistency. We hope to address some of the above issues in a future work.

\section*{Acknowledgments}

This work is supported by MINECO (Spain), FIS2016-76363-P (S.D.O).

\end{document}